%% file: multimodalRouting.tex
\pdfoutput=1

\documentclass[12pt, draftclsnofoot, onecolumn]{IEEEtran}

\usepackage[pdftex]{graphicx}
\usepackage[cmex10]{amsmath}
\usepackage{amssymb}
\usepackage{psfrag}
\usepackage{dsfont}
\usepackage[linesnumbered,ruled,vlined]{algorithm2e}
\usepackage{url}
\usepackage[table]{xcolor}
\usepackage{multirow}
\usepackage{wrapfig}
\usepackage{array,booktabs,arydshln}
\usepackage[caption=false]{subfig}
\usepackage{calc}

\newcommand{\vup}{\vspace{-1mm}}

\graphicspath{%
     {./figures/}%
}

\RequirePackage[normalem]{ulem} 
\RequirePackage{color}\definecolor{RED}{rgb}{1,0,0}\definecolor{BLUE}{rgb}{0,0,1} 

\newcommand{\PCYtilde}[1]{\widetilde{\cal Y}_{#1}}
\newcommand{\PCP}[2]{P_{#1,#2}}

\newcommand{\PCR}[4]{R_{#1}^{#4}(#2,#3)}
\newcommand{\PCRhat}[4]{\widehat{R}_{#1}^{#4}(#2,#3)}
\newcommand{\PCT}[3]{{\cal T}_{#1,#2}^{#3}}
\newcommand{\PCL}[2]{L_{#2}^{(#1)}}
\newcommand{\PCLtilde}[2]{\widetilde{L}_{#2}^{(#1)}}
\newcommand{\PCC}[3]{C_{#1,#2}^{(#3)}}
\newcommand{\PCY}[1]{{\cal Y}_{#1}}
\newcommand{\PCF}[2]{F_{#1}({#2})}
\newcommand{\PCdelta}[2]{\Delta_{#1}^{#2}}

\DeclareMathOperator*{\argmax}{arg\,max}

\newcolumntype{L}[1]{>{\raggedright\let\newline\\\arraybackslash\hspace{0pt}}m{#1}}
\newcolumntype{C}[1]{>{\centering\let\newline\\\arraybackslash\hspace{0pt}}m{#1}}
\newcolumntype{R}[1]{>{\raggedleft\let\newline\\\arraybackslash\hspace{0pt}}m{#1}}

\begin{document}
\title{Fair and Throughput-Optimal Routing\\[-3mm] in Multi-Modal Underwater Networks\vspace{0mm}}
\author{%
    \IEEEauthorblockN{ Roee Diamant$^{\S\ddag\star}$, Paolo Casari$^\sharp$, Filippo Campagnaro$^\S$,\\[-3mm]
                       Oleksiy Kebkal\raisebox{1ex}{\tiny $\maltese$}, Veronika Kebkal\raisebox{1ex}{\tiny $\maltese$}, Michele Zorzi$^\S$}\\
    \IEEEauthorblockA{\small $^\S$Department of Information Engineering, University of Padova, Italy\\[-2mm]}
    \IEEEauthorblockA{\small $^\ddag$Department of Marine Technology, University of Haifa, Israel\\[-2mm]}
    \IEEEauthorblockA{\small $^\sharp$IMDEA Networks Institute, Madrid, Spain\\[-2mm]}
    \IEEEauthorblockA{\small \raisebox{1ex}{\tiny $\maltese$}EvoLogics GmbH, Berlin, Germany\\[-2mm]}
    \IEEEauthorblockA{\small $^\star$Corresponding author, email: {\tt roeed@univ.haifa.ac.il}
    \vspace{-18mm}}
}
\date{}

\maketitle

\begin{abstract}
While acoustic communications have been considered the prominent technology to communicate under water for several years, other technologies are being developed based, e.g., on optical and radio-frequency electro-magnetic waves. Each technology has its own advantages and drawbacks: for example, acoustic signals achieve long communication ranges at order-of-kbit/s bit rate, whereas optical signals offer order-of-Mbit/s transmission rates but only over short transmitter--receiver distances. Such a technological diversity can be leveraged by multi-modal systems, which integrate different technologies and provide intelligence to decide which one should be used at any given time. In this paper, we address a fundamental part of this intelligence by proposing a novel routing protocol for networks of multi-modal nodes. The protocol makes distributed decisions about the flow in each link and over each technology at any given time, in order to advance a packet towards its destination. Our routing protocol prevents bottlenecks and allocates resources fairly to different nodes. We analyze the performance of our protocol via simulations and in a field experiment. The results show that our protocol successfully leverages all technologies to deliver data, even in the presence of imperfect topology information. To permit the reproduction of our results, we share our simulation code.
\end{abstract}


\maketitle



\input{intro.tex}


\input{relwork.tex}


\input{network_model_and_routing.tex}


\input{simulations.tex}


\input{experiment.tex}


\input{conclusions.tex}


\appendix

\input{appendix_disjoint_routes.tex}

\IEEEtriggeratref{1}
\IEEEtriggercmd{\renewcommand{\baselinestretch}{1.67}\footnotesize}


\input{acks.tex}

\bibliographystyle{IEEEtran}
\bibliography{IEEEabrv,refen,misc,biblio_Thesis_FC}

\end{document}

%% file: intro.tex

\section{Introduction} 
\label{sec:intro}

Several different physical layer (PHY) technologies have been developed to communicate under water. While most of them rely on acoustic communications at different frequencies and over different bandwidths~\cite{mandarreview-10}, optical communications are also gaining momentum~\cite{farr_integrated_OCEANS_2010,moriconi_hybrid_acoustic_optic_swarms_2015,bluecomm}, and recent work suggests that radio-frequency (RF) electro-magnetic communications are also finding their way into research interests~\cite{reevaluation} and system development~\cite{seatooth}. Novel system architectures based on electrostatic fields~\cite{electric_field_comms_wuwnet_2014} and magneto-inductive communications~\cite{multicoil_magneto_inductive_transceiver_UCOMMS_2012} are also being explored, albeit these technologies are still in their infancy compared to acoustics, optics, and electromagnetics.

Each of the above underwater PHY technologies offers a different balance of advantages and disadvantaged. The most prominent differences between underwater acoustic and optical communications, for instance, concern the data rate and the communications range. Acoustics typically provides low (order-of-kbit/s) transmission rates, but can cover ranges up to several~km. However, the performance of underwater acoustics is highly influenced by the environment, especially 
by strong and time-varying multipath. Underwater optical communication, on the other hand, provides a very high bandwidth on the order of up to several Mbit/s, but requires the transmitter and receiver to be very close, typically up to a few~m apart, and their transceivers to be aligned within each other's field of view. In addition, optical communications are sensitive to turbidity and tend to work best in dark waters. By way of contrast, RF communications do not need any alignment and can be developed based on very standard hardware already used for terrestrial radio systems; however, the conductivity of ocean waters attenuates RF waves within very short distances, and limits the achievable bit rates to less than 100~kbit/s within a distance of a few tens of~m~\cite{reevaluation}.

The above analysis suggests that there is a lot of potential in the integration of different PHYs into a multi-modal communication system. Such a system may be able to exploit the advantages of different technologies by transmitting through the best available one at any given moment. This approach was proposed, e.g., in~\cite{campagnaro_wireless_ROV_asilomar14}. After comparing the declared performance of the technologies available at the time, the authors concluded that a system encompassing optical and acoustic communications would be a good candidate for the wireless control of remotely operated vehicles. Notably, recent work~\cite{dol_moore_law_underwater_oceans_2015} supports the vision of multi-modal systems by showing that embedded processing platforms have sufficiently evolved to host the signal processing algorithms of acoustic communication systems on general-purpose computing platforms. This means that both the complexity and the versatility of these systems is already borne easily by current hardware, making multi-modal communications \emph{de facto} possible already with current technology.

A key role, in multi-modal communication systems, is played by the logic that decides how to switch between the available PHYs. While multi-modal point-to-point links are manageable with relatively simple policies~\cite{oceans2015genova_multimodal}, organizing multi-modal nodes into a network requires a complete change of perspective. In fact, the nodes may connect to different neighbors using (possibly partially overlapping) subsets of their PHYs. These subsets may change over time according to a variety of circumstances, that depend, e.g., on environmental conditions, mobility, and on the traffic requirements of the nodes. In this paper, we design a specific component of the multi-modal PHY usage logic: the multihop routing algorithm. We aim to provide a routing solution that fully utilizes the available PHY technologies in an optimized fashion. Specifically, by considering the different PHY technologies as another layer of network resources, we formalize the routing problem as a maximization problem where each node tries to extract the most from all its available PHYs. The solution to this problem leads to a routing protocol that is distributed and fair, and avoids bottlenecks. 
Our algorithm is valid in any network topology, and can be applied to any combination of available PHY technologies, including when different nodes incorporate different technologies.

Our contribution is twofold:
\begin{itemize}
    \item A novel distributed routing algorithm for multi-modal underwater networks, which maximizes the amount of information transmitted through all technologies available to each node, while at the same time balancing the traffic flow through the network and pursuing a fair network utilization for all nodes;
    \item A framework to handle both the cases where incomplete and complete topology information is available to each node.
\end{itemize}
We evaluate the performance of our routing algorithm by means of numerical simulations and in several field experiments performed in a lake north of Berlin with multi-modal nodes embedding different acoustic modems. To the best of our knowledge, this is the first reported trial for multi-modal routing schemes.

The remainder of this paper is organized as follows. Section~\ref{sec:relwork} discusses related work; Section~\ref{sec:routing} presents and formalizes our routing algorithm; Section~\ref{sec:sim} describes our simulation scenario and discusses the simulation results; Section~\ref{sec:trial} reports the results of the field experiments; Section~\ref{sec:concl} draws concluding remarks.

%% file: relwork.tex
\section{Related Work}
\label{sec:relwork}

The term ``multi-modal'' to refer to communication solutions encompassing diverse PHY subsystems is relatively new. However, an implementation of a multi-modal system with straightforward switching policies was already presented in~\cite{optics_modem_Vasilescu}, where the authors designed a data mule AUV that should approach each node of a deployed underwater acoustic network and retrieve data using optical communications. 
Acoustic and optical communications are typical technologies employed in multi-modal systems. As shown by the survey in~\cite{campagnaro_wireless_ROV_asilomar14}, radio frequency technologies for underwater communications are also under development, but their declared performance is topped by that of optical and acoustic systems at all distances of interest for RF technologies. In particular, the considerable maturity achieved by acoustic systems yields bit rates on the order of tens of kbit/s over fairly long distances~\cite{stojanovic2013recent}, and the higher than Mbit/s rates of optical modems in sufficiently benign waters are still unrivaled~\cite{farr_integrated_OCEANS_2010}. A notable feature of multi-modal systems is that the composition of multiple powerful PHY may not be necessary to achieve good performance. Indeed~\cite{lowcost_optical_modem_added_to_acoustic_UW_2014} demonstrates that even a low-bit rate, minimal-cost infrared optical modem, assembled starting from very inexpensive parts, can substantially improve the performance of underwater acoustic networks. The authors employed time synchronization and TCP connections as use cases.

The variable-depth moored nodes presented in~\cite{detweiler_acoustic_radio_depth_adj_2012} join acoustic communications under water and radio communications on top of the water surface. The system automatically finds a balance between the energy required for the node to reach the surface and employ radio communications, and the energy consumption of underwater acoustics, and chooses either strategy depending on the considered policy and on data transmission requirements. The autonomous underwater exploration platforms discussed in~\cite{kastner_multimodal_2012} is multi-modal in the sense that it can rely on multiple sensors, and has different underwater communication capabilities. Specifically, the authors discuss the tradeoffs between frequency-shift keying (FSK)-based modem technology and custom low-cost modems designed around a commercial off-the-shelf ceramic transducer~\cite{ucsd_lowcost}.
The work in~\cite{MURAO_acoustic_optic_2012} employs multi-modal optical and acoustic communications in a clustered optical underwater network. Specifically, the long range characterizing acoustic communications is exploited for cluster formation and management, whereas intra-cluster communications take place through optical connections. Q-learning~\cite{Watkins1992} is employed to set up and iteratively improve the routing structure in the network. 

Hybrid acoustic/optical multi-modal networks are considered for the transmission of real-time video streams in~\cite{youngtae_real_time_video_WUWNET_2014}. Bulk data streaming takes place through the optical channel, whereas acoustic communications are leveraged to send acknowledgments, to transmit data while during the alignment of optical modems, and as a fallback solution in turbid waters. The hybrid solution is shown via simulations to outperform both optical and acoustic communications alone.
In~\cite{basagnietal_max_value_info_infocom_2014}, the authors assume that sensor data generated by underwater nodes loses value over time. The path of an autonomous underwater vehicle (AUV) is then optimized to maximize the value of the information from the sensors. Sensor-to-AUV data upload takes place through an optical connection, whereas the sensors notify the AUV of new data using control packets through acoustic connections. Simple context-based switching schemes are considered in~\cite{campagnaro_wireless_ROV_asilomar14} to manage multi-modal optical and acoustic communications for the remote control of a remotely operated vehicle (ROV). By tuning the parameters of the switching policies, it is shown how the ROV-controller link can benefit from each technology and how well the ROV reacts to the controller's commands. This work was extended with the design of proactive switching policies in~\cite{oceans2015genova_multimodal}. In~\cite{oceans2016shanghai_multimodal}, more complex scenarios are implemented using the free-access DESERT underwater framework~\cite{opensourceDESERTandWOSS2014} and evaluated in a diver cooperation scenario. 

The authors in~\cite{moriconi_hybrid_acoustic_optic_swarms_2015} propose a hybrid acoustic/optical communications to coordinate swarms of autonomous underwater vehicles and to transfer information among swarm components. The custom design of both the acoustic and the optical modem is also discussed. A bilingual modem concept was implemented in~\cite{petroccia_bilingual_2015} using a custom re-configurable underwater acoustic modem. Two modulation schemes were employed for this purpose, namely the NATO standard JANUS and a higher-rate modulation format based on multi-level FSK. JANUS was employed both as a first-contact scheme and as a robust fallback scheme for harsh channel conditions, whereas the native FSK scheme was switched to upon first contact if channel conditions so allowed.

While the above works were key to introduce and improve the use of multi-modal technologies in underwater networks, the approaches taken are not formal and encompass heuristic solutions to divide the data between the PHY technologies. As a result, the multi-modal network is not optimally utilized. Considering this challenge, in this paper we present a routing algorithm that aims at maximizing the multi-modal network goodput under constraints that divide resources fairly among the nodes and avoid the generation of bottleneck nodes.

%% file: network_model_and_routing.tex
\section{Network model and Optimal Routing}
\label{sec:routing}

\begin{figure}
    \centering
    \subfloat[Flooding]{\includegraphics[width=0.225\textwidth]{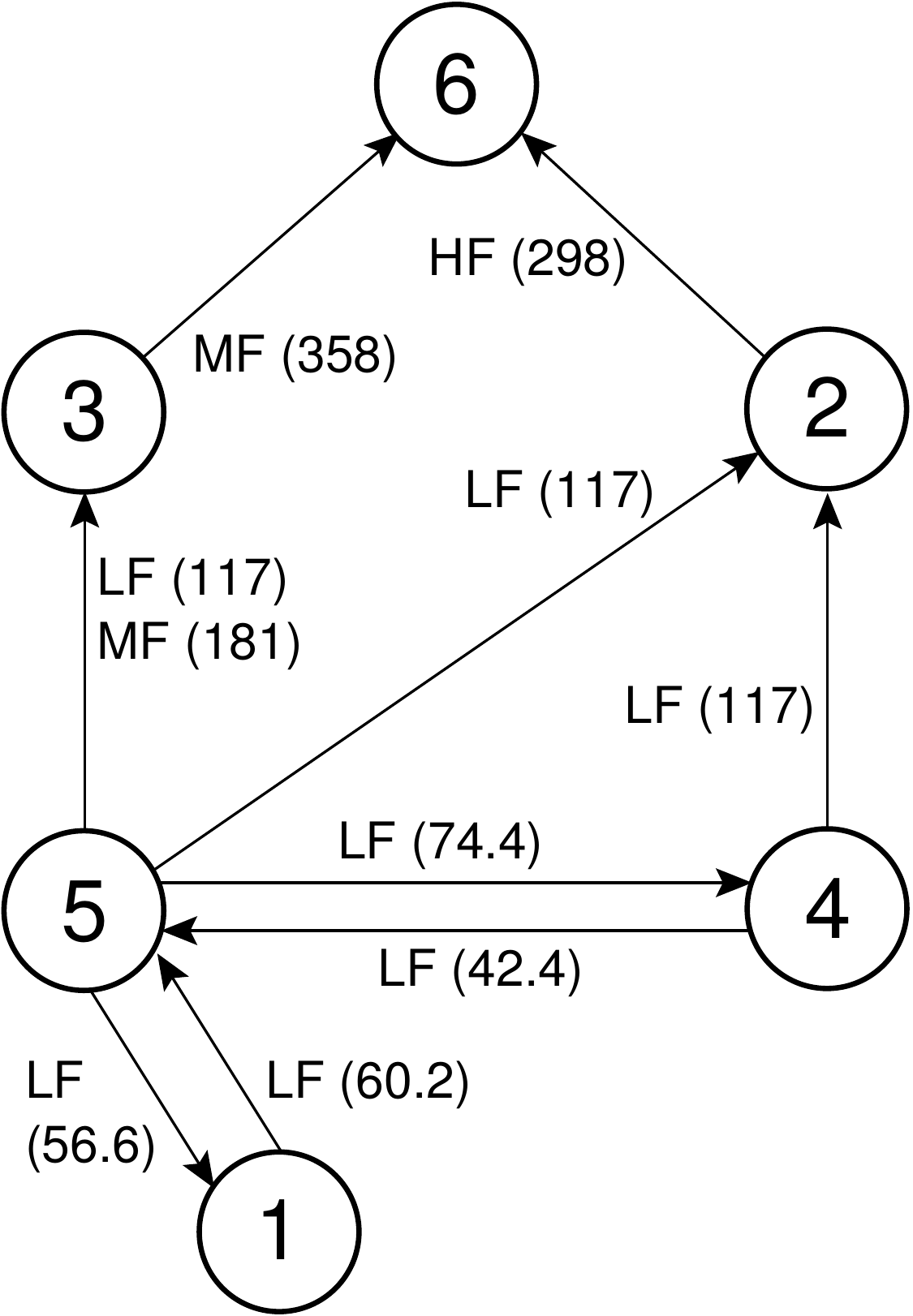}   \label{fig:keyidea.flood}}
    \hspace{7.5mm}
    \subfloat[OMR-PF]{\includegraphics[width=0.225\textwidth]{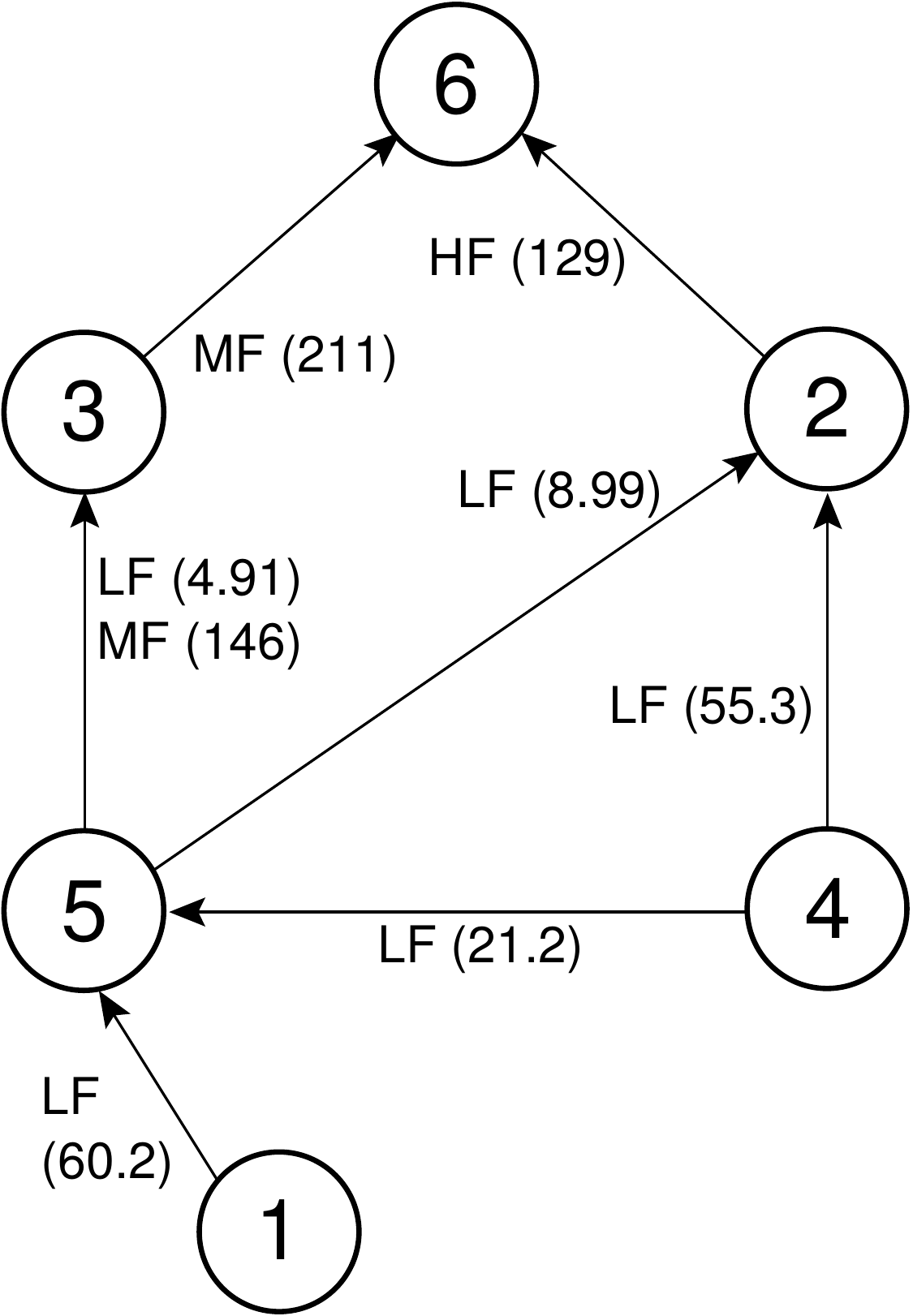}   \label{fig:keyidea.notopo}}
    \hspace{7.5mm}
    \subfloat[OMR-FF]{\includegraphics[width=0.225\textwidth]{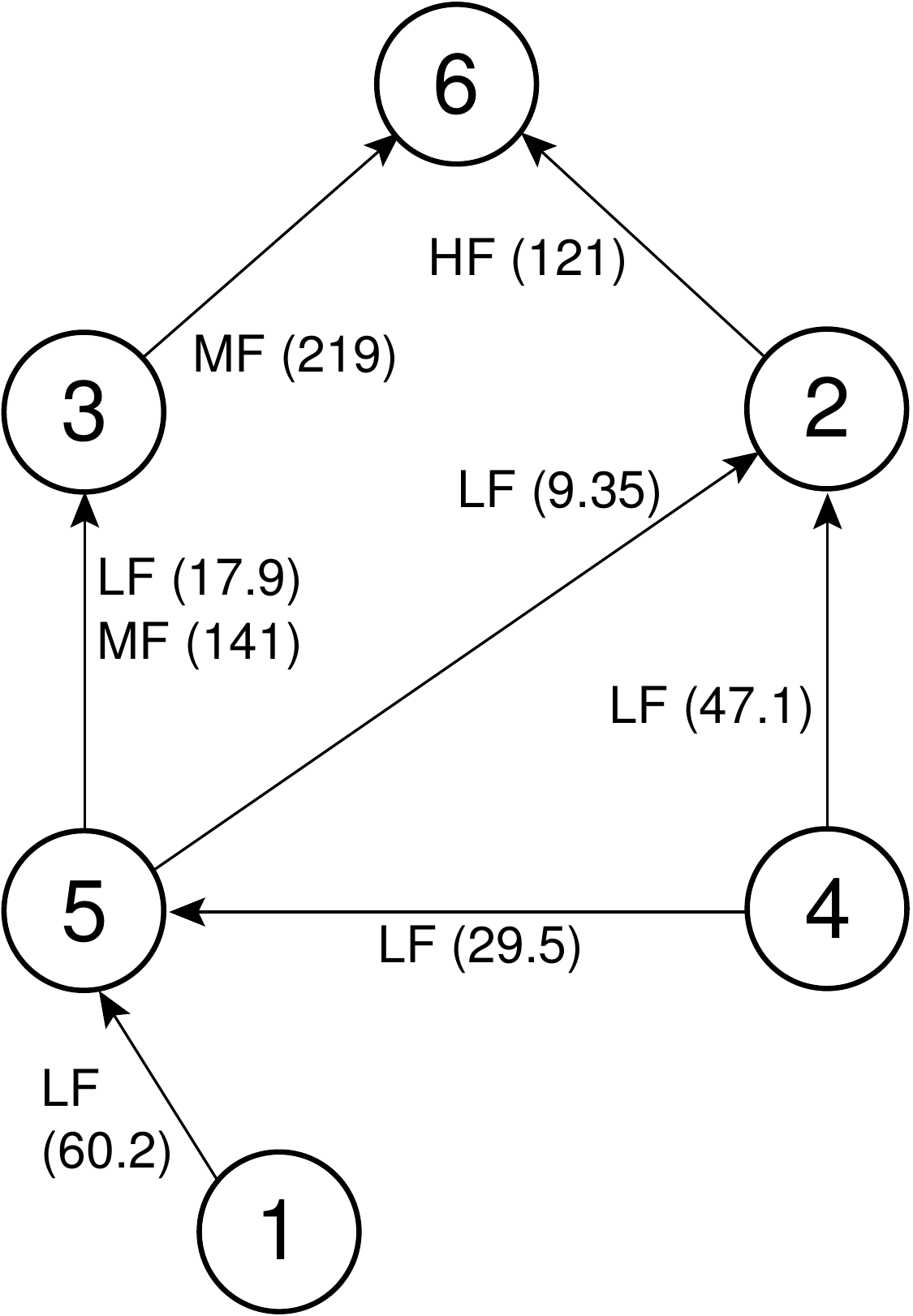}   \label{fig:keyidea.withtopo}}
    \caption{Link throughput in kbit/s of network operation using flooding (a) and our routing algorithm. For the latter, we consider a limited topology-informed (b) and a full topology-informed fairness computation (c).}\vspace{-5mm}
    \label{fig:keyidea}
\end{figure}

We consider a converge-casting network of multi-modal underwater nodes, where data traffic has to be routed towards a common sink node. We desire to obtain good performance in all key aspects of the multi-modal network. In particular, we are interested in minimizing the end-to-end transmission delay while maximizing the network goodput. Yet, since underwater networks usually face energy limitations, we are also interested in minimizing the packet transmission overhead. On the other hand, to keep network traffic flow smoothly and reduce bottlenecks, we are interested in the full exploitation of the available network links. 
With the goal of obtaining a favorable tradeoff between the above quality measures, in this paper we propose the first optimal multi-modal routing (OMR) scheme for underwater networks.

\subsection{Key idea}
\label{sec:routing.key}

The key idea behind our distributed routing scheme is that the available multi-modal links should be fully exploited, while at the same time avoiding that some relays become bottlenecks for the routing process. To do so, the nodes should $i$) avoid forwarding an excessive amount of traffic towards the relays upstream and $ii$) favor nodes with fewer valid routes to the sink during the data relaying process.
We achieve this by having the nodes estimate the capability of their relays to forward traffic further, and by having these relays distribute a minimal amount of information about the current backlog of data bits in their queues. This allows each network node to separately solve an optimization problem, and to find the number of bits to be transmitted to its neighbors through each multi-modal link. We note that our approach does not resort to flooding, as we explicitly want to avoid unnecessary redundant transmissions of the same data.

Even without topology information, this approach balances traffic much better with respect to a baseline algorithm that, e.g., floods all data through all available technologies. This can be observed in Fig.~\ref{fig:keyidea}, where we report the link throughput (expressed as the number of data bits correctly received per second of network operation) for a 6-node topology. The figure shows results for three multi-modal routing solutions: flooding where all links and all available technologies are used and packets are re-transmitted; a version of our OMR method to achieve fairness with only partial (one-hop) topology information (OMR-PF); and a version of our OMR method to achieve fairness with full topology information (OMR-FF). Our OMR-PF is a fully distributed protocol, while OMR-FF is centralized. Three acoustic PHY technologies are used: a low-frequency, low-rate technology (LF), a faster technology working at intermediate frequency (MF) and a high-frequency high-rate technology (HF). In parentheses, we show the obtained goodput in kbit/s. 

From Fig.~\ref{fig:keyidea}, we observe that, with respect to the flooding case~(Fig.~\ref{fig:keyidea.flood}), the link throughput values are much more balanced with our approach (Fig.~\ref{fig:keyidea.notopo}/\subref*{fig:keyidea.withtopo}). In the ideal case of OMR-FF where the full topology information is available (e.g., because the topology has been tested in advance, and does not change over time), the algorithm can better balance transmissions across all links compared to OMR-PF, as expected. For example, OMR-FF redistributes part of the traffic of node 4 through node 5, resulting in better utilization of the LF link from node~3 to node~6, as can be seen by comparing Figs.~\ref{fig:keyidea.notopo} and~\ref{fig:keyidea.withtopo}. 
In the following, we describe the routing algorithm in detail.

\subsection{Preliminary definitions and assumptions}
\label{sec:routing.defs}

\begin{table}
\centering
    \caption{Explanation of the employed notation}\vspace{-3mm}
    \label{tab:notation}
\resizebox{0.8\textwidth}{!}{%
    \renewcommand{\arraystretch}{1.4}
    \renewcommand{\baselinestretch}{0.95}\small
    \begin{tabular}{@{}L{1.5cm}L{7.5cm}L{4cm}L{2cm}@{}}
        {\bf Symbol} & {\bf Meaning} & {\bf Requires} & {\bf Shared with}\\
        \midrule
        ${\cal N}$ & Set of the network nodes & --- & --- \\
        $D$ & Destination node & Known by all nodes & --- \\
        $\PCY{i}$ & Set of upstream neighbors of node $i$ & --- & One-hop neighbors \\
        $\PCYtilde{i}$ & Set of all one-hop neighbors of node $i$ & --- & One-hop neighbors \\
        ${\cal T}_i$ & Set of PHY technologies available at node $i$ & --- & --- \\
        $\PCT{i}{j}{\tau}$ & Set of PHY technologies through which $i$ can transmit to $j$ at time $\tau$ & Technology availability signaled by MAC protocol & --- \\
        $\PCP{j}{\tau}$ & Number of bits in node $j$'s queue at time $\tau$ &  $\PCP{j}{\tau'}$ for $\tau' < \tau$, $\PCRhat{j}{k}{t}{\tau}$ $\forall k \in \PCY{j}$ & One-hop neighbors \\
        $\PCC{i}{t}{u}$ & Total number of bits that can be transmitted by $i$ using technology $t$ over a time period of duration $u$ & Technology availability signaled by MAC protocol & --- \\
        $\PCL{i}{j}$ & Number of node-disjoint routes towards the destination $D$ available to $i$ when routing through $j$ & Topology information or $\PCY{\ell} \ \forall \ell \in \PCY{i} \cup \PCYtilde{i}$ & --- \\
        $F_j(i)$ & Fair share of $j$'s upstream transmission resources that can be dedicated to node $i$ & $\PCL{i}{j}$, $\PCY{\ell} \ \forall \ell \in \PCY{i} \cup \PCYtilde{i}$ & --- \\
        $\PCR{i}{j}{t}{\tau}$ & Number of bits in node $i$'s queue sent to node $j$ using technology $t$ at time $\tau$ & $\PCP{i}{\tau}$, $\PCdelta{j}{\tau}$, $\PCC{i}{\tau}{u}$, $F_j(i)$ & --- \\
        $\PCdelta{j}{\tau}$ & Amount of upstream transmission resources of node $j$ that can be secured for node $i$'s transmissions  & $\PCC{j}{t}{u} \forall t \in \PCT{j}{k}{\tau}$, $\PCR{j}{k}{t}{\tau}$ $\forall k \in \PCY{j}$& Nodes $\ell$ s.t.\ $i \in \PCY{\ell}$ \\
        \midrule
    \end{tabular}%
}
            \vspace{-5mm}
\end{table}

We assume that our underwater network is composed of a set ${\cal N}$ of multi-modal nodes, where $|{\cal N}| = N$. The network implements a converge-casting scenario, where all nodes send their information to a common sink (denoted as $D$, e.g., node~6 in Fig.~\ref{fig:keyidea}) over multiple hops.
We assume that the network topology has been already discovered.
    \footnote{This can be done, e.g., by sending beacon packets downstream from the sink to the network nodes~\cite{iris_wsn_rossi_2013}, or by carrying out specific processes aimed at discovering either the topology itself~\cite{ted_roee_2016} or at least the available routes~\cite{sun-evologics-2012}. The discovered structure can be maintained by tracking transmissions successes over each link over time~\cite{CARP_journal_2015}. These processes are out of the scope of this work.}  
While the implementation of a topology discovery algorithm is outside the scope of this paper, we nonetheless do assume that the process can be subject to errors, or to inaccuracies due to slow topology changes over time. We will take these errors into account in the design of the routing protocol.
Given the outcome of the routing structure discovery, we assume that each node knows the alternatives he has to forward a packet towards the sink $D$. Accordingly, for each node, we call $\PCY{i}$ the set of upstream neighbors of $i$, i.e., $\PCY{i}$ contains all one-hop neighbors of $i$ that can advance packets one further hop towards $D$ (for example, $\PCY{5} = \{2,3\}$ in Fig.~\ref{fig:keyidea}).%
    \footnote{We remark that routing in the network is never performed downstream, i.e., no relaying operation will bring a packet one hop farther from the destination.}   %
We also call $\PCYtilde{i}$ the list of \emph{all} one-hop neighbors of $i$.

Each multi-modal node incorporates a number of PHY technologies, listed in the set ${\cal T}_i$ (e.g. ${\cal T}_5 = \{\textrm{LF}, \textrm{MF}\}$ in Fig.~\ref{fig:keyidea}). The nodes can communicate using any technology simultaneously available to it and to the addressed receiver. Note that the list of available technologies may vary over time, e.g., due to channel variations or mobility. Let $\PCT{i}{j}{\tau}$ be the set of technologies that $i$ can use to transmit to $j$ at time $\tau$. We assume that this set of technologies is known to the routing protocol, e.g., because some underlying MAC protocol forwards a notification when a given technology is available. This process is outside the scope of this paper, and can be implemented, e.g., through the schemes in~\cite{oceans2015genova_multimodal,oceans2016shanghai_multimodal}.

Each node maintains a queue with a list of packets to transmit. Denote the bits in node $i$'s queue at time $\tau$ as $\PCP{i}{\tau}$. Define $\PCR{i}{j}{t}{\tau}$ as the number of bits in $\PCP{i}{\tau}$ that will be sent by node $i$ to node $j \in {\cal Y}_i$ using technology $t$ at time $\tau$. The objective of the routing algorithm is to find optimal values for $\PCR{i}{j}{t}{\tau}$, under a constraint on the total number of bits that can be transmitted by $i$ using technology $t$ over a time span $u$, denoted as $\PCC{i}{t}{u}$. We will indicate these optimal values as $\PCRhat{i}{j}{t}{\tau}$.
A summary of the employed notation is provided in Table~\ref{tab:notation}. The table also reports the inter-dependencies among the quantities introduced above, and the nodes each quantity is shared with. This is meant as a reference for the algorithm description below.

The optimization is to be carried out using the information available at node $i$ or passed on by its upstream neighbors. In particular, we assume that node $i$ knows: 
$\PCY{i}$ and $\PCY{j}$ $\forall j \in \PCY{i}$; $\PCP{i}{\tau}$ and $\PCP{j}{\tau'}$ $\forall j \in \PCY{i}$, where $\tau' < \tau$ is a time epoch that refers to a transmission carried out by node $j$ immediately preceding the current epoch $\tau$; and
$\PCC{j}{t}{u}$ $\forall j \in \PCY{i}$.

\subsection{Routing algorithm} \label{sec:routing.algo}
We are now ready to describe the steps of the routing optimization algorithm executed by node $i \in {\cal N}$. Node $i$ has to decide how many bits to transmit through each of its available technologies, and carries out the following steps for each upstream neighbor in $\PCY{i}$. For clarity, we will illustrate the algorithm by referring to one of these upstream nodes, $j$.
%
The optimal transmitted bit allocation for node $i$ is obtained by solving the following problem:
\begin{subequations}
\label{e:rxi}
    \begin{align}
        \PCRhat{i}{j}{t}{\tau} = & \argmax_{\PCR{i}{j}{t}{\tau}}
             \sum_{j \in \PCY{i}} \; \sum _{t \in \PCT{i}{j}{\tau}}  \PCR{i}{j}{t}{\tau}  \label{eq:rxi} \\
        \textrm{s.t.} \phantom{=} & \sum_{j \in \PCY{i}} \; \sum_{t \in \PCT{i}{j}{\tau}} \PCR{i}{j}{t}{\tau} \leq \PCP{i}{\tau} \, ; \label{eq:rxi_c1} \\
                                  & \sum_{t \in \PCT{i}{j}{\tau}} \PCR{i}{j}{t}{\tau} \leq \PCdelta{j}{\tau} \, ; \label{eq:rxi_c2} \\
                                  & \PCR{i}{j}{t}{\tau} \leq \PCC{i}{t}{u} \PCF{j}{i} \, .  \label{eq:rxi_c3}
    \end{align}
\end{subequations}
Constraint~\eqref{eq:rxi_c1} means that the bits transmitted across all technologies shall not exceed the remaining number of bits in queue at node $i$. Constraint~\eqref{eq:rxi_c2} takes into account that $i$'s upstream neighbor $j$ may have a backlog of packets to be transmitted, and that node $j$ would give priority to these bits in a FIFO fashion. Assuming that the remaining portion of $j$'s upstream transmission resources after the transmission of the backlog is sufficient to transmit $\PCdelta{j}{\tau}$ bits from node $i$, constraint~\eqref{eq:rxi_c2} makes sure that $i$ transmits no more than $\PCdelta{j}{\tau}$ bits to $j$, aggregate over all technologies. Finally, constraint~\eqref{eq:rxi_c3} means that the number of bits transmitted through either technology should not exceed a certain limit, defined as $i$'s fair share of $j$'s upstream link capacity, where $\PCF{\ell}{j} = 0$ if node $\ell$ has nothing to transmit.  

Note that $\sum_{i}\PCF{j}{i}$ can exceed 1. This is because condition \eqref{eq:rxi_c3} only applies if a node $i$ has more possible relays to the sink than node $j$: in this case, it should divide its transmissions while considering the relay options of other neighbors of $j$. Also note that since we limit ourselves to a distributed solution, node $i$ has typically no way to ascertain the technology used over link $j \rightarrow k$, $k \in \PCY{j}$. Instead, we perform technology allocation only hop-by-hop. As a result, the term $\PCF{j}{i}$ is not related to the used technology $t$. 

The quantities required to evaluate the constraints are fully determined by node $i$. Node $i$ is assumed to know the capacity of its one-hop links, its available technologies, and its different paths to the sink. However, $\PCdelta{j}{\tau}$ and $F_j(i)$ must still be computed, as will be detailed in the following.


\subsubsection{Calculation of the fair share of node $j$'s resources}\label{sec:fair}
$\\$
We start with the computation of $F_j(i)$. 
The upstream transmission resources of node $j$ are assigned to a downstream neighbor $i$ depending on the number of node-disjoint routes towards the destination $D$ available to $i$, indicated with $\PCL{i}{j}$, where the subscript $j$ indicates that $j \in \PCY{i}$, and that it is being considered as a next hop. 
The rationale behind the resource assignment strategy is that if some downstream neighbors $m$ of $j$ can reach the destination only via a route that passes through node $j$, such nodes $m$ should be given a higher priority in the use of $j$'s upstream transmission resources.
Formally, define 
\begin{equation}
\label{e:L}
\PCLtilde{i}{j} = \sum_{\ell \in \PCY{j}} \PCL{\ell}{j} - \PCL{i}{j}\;.
\end{equation}
If $\PCLtilde{i}{j} = 0$, then we immediately set $F_j(i)=1$, as $j$ is the only neighbor of $i$ that can relay packets towards $D$ (e.g., $F_5(1)=1$ in Fig.~\ref{fig:keyidea}). Otherwise, $F_j(i)$ is computed as
\begin{equation}
    F_j(i) = \frac{ \PCLtilde{i}{j} }{ \sum_{\ell \in \PCY{j}} \PCLtilde{\ell}{j} }  \, ,
\end{equation}
where it is understood that $i \notin \PCY{j}$, i.e., $i$ is not an upstream neighbor of $j$. Note that this is a way of ``fairly'' allotting more resources to nodes with fewer available routes, not a means to split the capacity of node $j$'s links towards its upstream neighbors, which is instead taken care of distributely via constraint~\eqref{eq:rxi_c2}. Instead, we allow nodes with a single forwarding opportunity to convey all traffic there, while nodes with additional opportunities should split their traffic through all available routes. For example, in Fig.~\ref{fig:keyidea}, node~1 can only forward to node~5, so $\PCF{5}{1}=1$, and because of constraint~\eqref{eq:rxi_c3}, all of node~1's traffic will be conveyed through the link 1$\to$5, which can transport $\PCC{5}{\rm LF}{u}$~bits over a time span $u$. Conversely, node~4 shares node~5 as a potential relay, but has an additional opportunity to forward to node~2: for this reason, $\PCF{4}{5}= 1/3$ and $\PCF{4}{5}= 2/3$, hence node~4 will send up to $\PCC{4}{\rm LF}{u} \PCF{5}{4}$ to node~5 and $\PCC{4}{\rm LF}{u} \PCF{5}{2}$ to node~2, as per constraint~\eqref{eq:rxi_c3}.

$\PCL{i}{j}$ in \eqref{e:L} is computed differently depending on the network topology information available to node $i$. We hereby distinguish between two cases: a) full topology-informed fair share computation, in case perfect topology information is available to $i$; and b) one-hop topology-informed fair share computation, otherwise.

In case of OMR-FF (case a), we assume that node $i$ is aware of the full network graph, which makes it possible for the node to exactly compute the number of disjoint routes available to itself and its neighbors, thereby exactly computing its own fair share of resources. The algorithm we employ to do so is provided in the Appendix.

In case of OMR-PF (case b), only one-hop topology information is available to $i$. In this case, $\PCL{i}{\tau'}$ is estimated  as
\begin{equation}
    \PCL{i}{j} = |\PCY{j}| - \sum_{ w \in \PCYtilde{i} \cup \PCY{j} } \mathds{1} \big[ \PCY{w}=\{i,j\} \vee \PCY{w}=\{i\} \vee \PCY{w}=\{j\} \big]
    \label{eq:Lkj_onehopinfo}
\end{equation}
where $\mathds{1}[{\sf p}]$ evaluates to $1$ whenever the predicate ${\sf p}$ is true. Eq.~\eqref{eq:Lkj_onehopinfo} means that, as a best effort, $\PCL{i}{j}$ is assumed to be equal to the number of downstream neighbors of $k$, decreased by one for each node $w$ that has only $i$, $j$, or both as upstream neighbors, which may occur due to erroneous topology information. Without such reduction, node $i$ would be given an excessive resource share. We note that the computation of $\PCL{i}{j}$ in \eqref{eq:Lkj_onehopinfo} is not carried out if the destination $D \in \PCY{j}$. In this case, the traffic is always directed to the sink, without passing through other 1-hop neighbors.

\subsubsection{Calculation of upstream resources}
$\\$
We proceed with the computation of $\PCdelta{j}{\tau}$ from \eqref{eq:rxi_c2}, which represents the amount of $j$'s upstream transmission resources that can be assigned to node $i$. This computation is made based on an estimate (obtained by $i$) of the quantities $\PCRhat{j}{k}{t}{\tau}$ $\forall k \in \PCY{j}$. These quantities estimate the outcome of the allocation problem as may be solved by $j$ to compute how many bits it should transmit to its own upstream neighbors with each technology $t \in \PCT{j}{k}{\tau}$. We have 
\begin{equation}
    \PCdelta{j}{\tau} = \sum_{k \in \PCY{j}} \; \sum_{t \in \PCT{j}{k}{\tau}} \bigg(  \PCC{j}{t}{u} - \PCRhat{j}{k}{t}{\tau} \bigg)\;.
    \label{eq:deltaflow}
\end{equation}
Note that we still indicate the current time $\tau$ as a reminder that the current solution to $i$'s problem depends on $j$'s solution for its current transmission allocation. The quantities $\PCRhat{j}{k}{t}{\tau}$ are obtained by $i$ by solving the following problem:
\begin{subequations}
\label{e:rxj}
    \begin{align}
        \PCRhat{j}{k}{t}{\tau} = & \argmax_{\PCR{j}{k}{t}{\tau}} 
            \sum_{k \in \PCY{i}} \; \sum _{t \in \PCT{j}{k}{\tau}(k,t)}  \PCR{j}{k}{t}{\tau}  \label{eq:rxj} \\
        \textrm{s.t.} \phantom{=} & \sum_{k \in \PCY{j}} \; \sum_{t \in \PCT{j}{k}{\tau}} \PCR{j}{k}{t}{\tau} \leq \PCP{j}{\tau} \, ; \label{eq:rxj_c1} \\
                                  & \PCR{j}{k}{t}{\tau} \leq \PCC{j}{t}{u} \, \PCF{k}{j} \, \label{eq:rxj_c2} ,
    \end{align}
\end{subequations}
where $F_k(j)$ is the share of node $k$'s resources that can be devoted to transport node $j$'s traffic.
Constraint~\eqref{eq:rxj_c1} means that the bits transmitted through all technologies shall not exceed the remaining number of bits in node $j$'s queue, whereas constraint~\eqref{eq:rxj_c2} implies that the number of bits transmitted by $j$ via either technology shall not exceed its share of the upstream capacity of its relay $k$ over a time period of length $u$. No constraint is imposed based on the terms $\PCdelta{k}{\tau}$, as $i$ does not know them and it would take too many resources for $j$ to transmit the corresponding information, especially over slow acoustic links. Note that~\eqref{eq:deltaflow} enforces congestion control in the network, by avoiding that a downstream node transmits more data than the receiving relay can advance towards $D$. 

We note that the same procedure described in Section~\ref{sec:fair} above is employed to compute the fairness values $F_k(j)$ $\forall  k \in \PCY{j}$. In case of OMR-FF (full topology) this procedure is trivial. However, for case OMR-PF, node $i$ cannot calculate $F_k(j)$ without knowledge of $\PCY{k}$ $\forall  k \in \PCY{j}$, and therefore must rely on node $j$ to transmit the $F_k(j)$ values. Similarly, since node $i$ is not aware of $\PCP{j}{\tau}$, we let $j$ piggyback this value into each transmission. As a result, the overhead of this information is in the order of only a few bits. 
Assume that $j$ communicated $\PCP{j}{\tau'}$ at some preceding instant $\tau' < \tau$: $\PCP{j}{\tau}$ can be readily derived as
\begin{equation}
    \PCP{j}{\tau} = \PCP{j}{\tau'} - \sum_{k \in \PCY{j}} \; \sum_{t \in \PCT{j}{k}{\tau'}} \PCRhat{j}{k}{t}{\tau} \, .
    \label{eq:phat}
\end{equation}

\begin{wrapfigure}[6]{r}[0pt]{3.5in}
    \centering
    \vspace{-16mm}
    \includegraphics[width=0.48\textwidth]{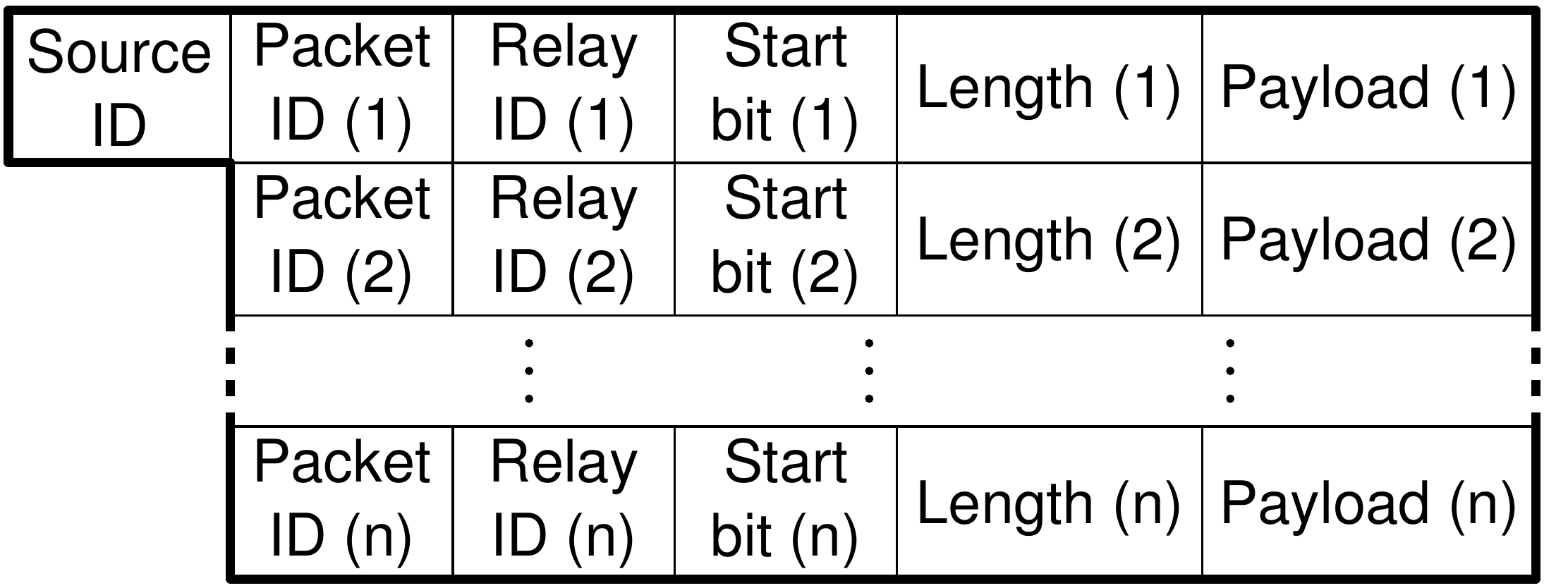}
    \caption{Format of a typical transmitted datagram formed by fragments taken from $n$ packets.}
    \label{fig:pktformat}
\end{wrapfigure}

The nodes decide on the number of bits to be transmitted over each link regardless of what packet they actually belong to, and manage the queue in a FIFO fashion. Therefore, the format of each transmitted datagram is such that the receiver can properly fragment and re-assemble the transmitted bits. A scheme of a typical transmitted datagram is provided in Fig.~\ref{fig:pktformat}.

\subsection{Complexity and overhead of OMR}

To obtain the routing solution with OMR, each node $i$ needs to solve \eqref{e:rxi} and \eqref{e:rxj}. Since both $\PCRhat{i}{j}{t}{\tau}$ and $\PCRhat{j}{k}{t}{\tau}$ can take any value, these two optimization problems are solved through linear programming. The average complexity of OMR-PF is therefore polynomial with $|\PCY{i}|\cdot\underset{j}{\max}|\PCT{i}{j}{\tau}|$. OMR-FF requires the additional computation of disjoint routes (see Algorithm~\ref{Algo1} in the Appendix) whose complexity is ${\cal O}\left(N^2\right)$.

In terms of overhead, OMR-FF requires full knowledge of the network topology, which translates into the transmission of $N^2$~bits every time a single link changes in the network. The discovery of the network topology is performed by multiple packet transmissions with a per-node overhead of $\log(N)$~bits re-transmitted roughly $N$ times by methods such as, e.g.,~\cite{ted_roee_2016}. OMR-FF also requires the transmission of the size of the queue of one-hop neighbors, $\PCP{j}{\tau}$. Representing $\PCP{j}{\tau}$ as one byte, the total overhead of OMR-FF is therefore $N^2+ N^2\log(N)+8N$~bits. Contrary to OMR-FF, OMR-PF is fully distributed and requires information only from one-hop links to transfer $\PCY{j}$, $F_k(j), \ \forall  k \in \PCY{j}$, and $\PCP{j}{\tau}$. This information is piggybacked to node $i$ within the packet transmitted by its one-hop neighbor $j$ only when the values change. Since the network topology changes very slowly, $\PCY{j}$ and $F_k(j)$ are rarely transmitted. However, $\PCP{j}{\tau}$ is transmitted after each packet transmission. The communication overhead of OMR-PF is therefore only $8N$~bits.

%% file: simulations.tex
\section{Simulations}\label{sec:sim}

We now explore the performance of our routing scheme, named optimal multi-modal routing (OMR), through numerical simulations. 
We consider both flavors of OMR defined in Section~\ref{sec:routing.key}, namely \textit{OMR--FF}, where fair shares of resources are calculated based on full topology information, and \textit{OMR--PF}, where the fair share computation is based only on the knowledge of one-hop links. 
We recall that this difference affects the way resources are allocated to different nodes over a multi-hop path. 
Namely, with only local topology information, OMR--PF is more conservative in terms of link capacity usage than OMR--FF. Except for this aspect, the two OMR flavors behave in the same way. 
For reproducibility, we publish the implementation of both versions of OMR.%
\footnote{The code is available for download at \url{http://marsci.haifa.ac.il/share/diamant/MultiModalRoutingCode.zip}~.}

With the absence of a benchmark routing scheme for multi-modal networks, we compare the performance of the two versions of OMR with that of a \textit{flooding} mechanism, in which a node broadcasts all incoming packets through all available technologies. To avoid loops in the flooding scheme, we include in each packet the routing path it has traveled. A receiver will then avoid broadcasting the packet if this routing path shows that the packet has already traveled through all its one-hop neighbors. In the flooding method, packets are fragmented according to the maximum length allowed by the technology through which the packet is sent.  

\subsection{Quality metrics}

We measure performance in terms of the end-to-end transmission delay, per-node goodput,  message success rate transmission efficiency, and link throughput. Once all fragments of a packet $i$ of node $n$ have been successfully received by the sink, we measure the message's end-to-end transmission delay as
\begin{equation}
    \rho_d=\frac{1}{N-1}\sum_{n=1}^{N-1}\frac{1}{R_n}\sum_{i=1}^{R_n} (T^r_{n,i}-T^s_{n,i}) \;,
    \label{e:delay}
\end{equation} 
where $T^{\mathrm{r}}_{n,i}$ is the time the full message was received, $T^{\mathrm{s}}_{n,i}$ is the time the message reached the network layer for routing, and $R_n$ is the number of messages sent by node $n$ and received in full by the sink node. For a network run time $T_{\rm net}$, the per-node goodput is defined by
\begin{equation}
    \rho_g=\frac{1}{N-1}\sum_{n=1}^{N-1}\sum_{i=1}^{I_n}\frac{M^{\mathrm{r}}_{n,i}}{T_{\rm net}}\;,
    \label{e:goodput}
\end{equation}
where $M^{\mathrm{r}}_{n,i}$ is the number of bytes received by the sink for a message $i$ originated from node $n$, and $I_n$ is the number of messages originated by node $n$. And the message success rate is
\begin{equation}
    \rho_s=\frac{1}{N-1}\sum_{n=1}^{N-1}\frac{R_n}{I_n}\;.
    \label{e:MER}
\end{equation}

Note that $M^{\mathrm{r}}_{n,i}$ from \eqref{e:goodput} can exceed the number of bytes transmitted by node $n$, denoted by $M^{\mathrm{s}}_{n,i}$. This case happens when message $i$ or parts of it are sent through several links such that the sink may receive multiple copies. We consider these cases a resource waste, and refer to this excess of copies as a transmission overhead. This overhead is measured by
\begin{equation}
    \rho_o=\frac{1}{N-1}\sum_{n=1}^{N-1}\sum_{i=1}^{I_n}U\left(\frac{M^r_{n,i}}{M^s_{n,i}}-1\right)\;,
    \label{e:overhead}
\end{equation}
where $U(x)$ is a step function whose value equals 1 if $x>0$, and zero otherwise. Another energy efficiency metric is the total number of transmitted bytes across the network for a single message. This is defined by
\begin{equation}
    \rho_e=\frac{1}{(N-1) \sum_{n=1}^{N-1} I_n} \sum_{n=1}^{N-1}\sum_{i=1}^{I_n}\frac{B_{i,n}}{T_{\rm net}}\;,
    \label{e:tx}
\end{equation}
where $B_{i,n}$ is the total number of bytes transmitted for message $i$ originated from node $n$. 

Finally, the throughput of the link from node $n$ to node $m$ using communication technology $t$ is defined as the ratio between the number of bytes successfully transmitted through the link, $R^{t}_{n,m}$, and the run time. Formally, the average link throughput is
\begin{equation}
    \rho_u=\frac{1}{N^t}\sum_{n\in{\cal N}^t}\frac{1}{D^t_n}\sum_{m\in{\cal D}^t_n}\frac{R^{t}_{n,m}}{T_{\rm net}}\;,
    \label{e:LU}
\end{equation}
where ${\cal N}^t$ is the set of the nodes who hold communication technology $t$, and ${\cal D}^t_n$ is the set of the nodes that share a communication link with node $n$ via technology $t$. Moreover, $|{\cal N}^t| = N^t$ and $|{\cal D}^t_n| = D^t_n$.

As mentioned in Section~\ref{sec:routing.key}, we desire to minimize $\rho_d$, and to maximize $\rho_g$ and $\rho_s$. Yet, for energy conservation, we are also interested in minimizing $\rho_o$ and $\rho_e$. Finally, for better fairness and to avoid congestion, we are interested in a large $\rho_u$.
 
\subsection{Simulation setup}\label{sec:sim.setup}

Our simulation setup is based on a Monte-Carlo set of~1000 network topologies. In each simulation run, $N=10$ nodes are placed uniformly at random over an area of 500$\times$500~m\textsuperscript{2} with water depth of 100~m. The line-of-sight between the nodes may be interrupted by four horizontal obstacles and one vertical obstacle at uniformly distributed locations with uniformly distributed length in the range $[10,50]$~m. Node 10 is defined as the sink node. Each of the other nine nodes is equipped with one or more communication technologies at random between low frequency acoustics, mid frequency acoustics, and high frequency acoustic communications. The characteristics of the three acoustic systems are based on the 18--34~kHz, the 48--78~kHz, and the 120--200~kHz EvoLogics\newpage

\begin{wraptable}[6]{r}[0pt]{3.5in}
        \centering
        \vspace{-3mm}
        \caption{Simulations: characteristics of the simulated communication technologies}\vup\vup
        \label{t:comm}
        \renewcommand{\arraystretch}{1.1}
        \renewcommand{\baselinestretch}{1.00}\footnotesize
        \begin{tabular}{@{\hspace{1mm}} lll @{}}
            Technology & Bit rate [bps] & Max range [m] \\
            \midrule
            Low-rate acoustics & 1000 & 3000 \\ 
            Mid-rate acoustics & 32000 & 300 \\ 
            High-rate acoustics & 64000 & 100 \\
            \midrule
        \end{tabular}
\end{wraptable}
\noindent modems~\cite{evologics_modems_page}. A summary of these characteristics is provided in Table~\ref{t:comm}, where the communications ranges of each model has been conservatively set.

We run each simulation for $T_{\rm net}=600$~s. At the beginning of each simulation, each of the nine nodes generates its own packets according to a Poisson process of rate $\lambda=3$ packets per minute per node. The size of each packet is drawn uniformly at random between~0 and 64~kbit. At any given time, the node is either idle, or serving a self-generated message or a packet received by another node. For each served packet, the node solves the routing allocation problem, as discussed in Section~\ref{sec:routing}. The packet is then segmented according to the solution of the routing problem and sent over the different links according to the determined routing allocation. Besides the information bearing Bytes, each packet segment includes the ID of the original message, the location of the packet segment within the original message, and the routing path the packet segment has gone through. Once received at the sink node, the various packet segments belonging to the same message are combined together. 

We consider a binary phase-shift-keying modulation, and a scheduling protocol where a node holding a packet transmits it as soon as all its communication technologies are free. Once a packet is received, an acknowledgment is transmitted. To form the full topology information required for the OMR-FF method and the one-hop link information required for OMR-PF, we refer to the communication ranges in Table~\ref{t:comm}. For example, for mid-frequency acoustic communications, a link would be assumed to exist if the distance between the two nodes is smaller than 300~m, and this distance is continuously measured in our simulations by an underlying PHY mechanism. To calculate the route on the way to the sink (i.e., the sets $\PCY{i}$, $\forall i$), we carry out a preliminary route discovery phase, where the sink propagates a discovery packet through the network. The discovered routes are kept stable throughout each simulation run. 

While in the OMR scheme the full topology (OMR--FF) or the one-hop links (OMR--PF) are assumed to be known, in reality links would vary from the communication range set in Table~\ref{t:comm}. To simulate this, we calculate the instantaneous packet error rate (PER) for each link used by transmitted packet segments. 
Once a packet fails and no acknowledgment is received, the packet is shifted to the end of the message queue and is re-transmitted at a later time. 
The PER is computed based on the simulated signal-to-noise ratio (SNR) and on the packet size. The SNR of the low-rate and high-rate acoustic links is calculated using the Bellhop framework~\cite[Ch.~3]{schmidt} for shallow waters of depth 100~m, flat sand bottom, fixed sound speed of 1500~m/s, and considering a source level of 170~dB re (1~$\mu$Pa at 1~m). The ambient noise level is set as 40~dB re (1$\mu$Pa\textsuperscript{2}/Hz) for the low-frequency acoustics, as 30~dB re (1$\mu$Pa\textsuperscript{2}/Hz)
for the mid-frequency acoustics, and 10~dB re (1$\mu$Pa\textsuperscript{2}/Hz) for the high-frequency acoustics.

We consider two MAC schemes. The first (\textit{Ideal}), is an ideal schedule where no packet collisions occur and acknowledgments are assumed to always arrive. This ideal channel works in favor of the flooding scheme where the links are expected to be utilized in full. The second (\textit{Immediate}) is a MAC protocol in which packets are transmitted immediately upon arriving to the MAC layer, unless another transmission or reception is already taking place, and the reception of packets and acknowledgments is determined based on the link SNR and only when no collision occurs with another packet or acknowledgment. The Immediate MAC is the same protocol employed by the modems in the lake experiment described in Section~\ref{sec:trial}~\cite{dmac-EvoLogics}. In both the Ideal and Immediate MAC approaches, packets that need to be re-transmitted are re-inserted as new packets at the end of the queue. 

\subsection{Simulation results}\label{sec:sim.results}

\begin{figure}[t]
    \centering
    \subfloat[Ideal MAC.\label{f:Delay}]{\includegraphics[width=0.4\textwidth]{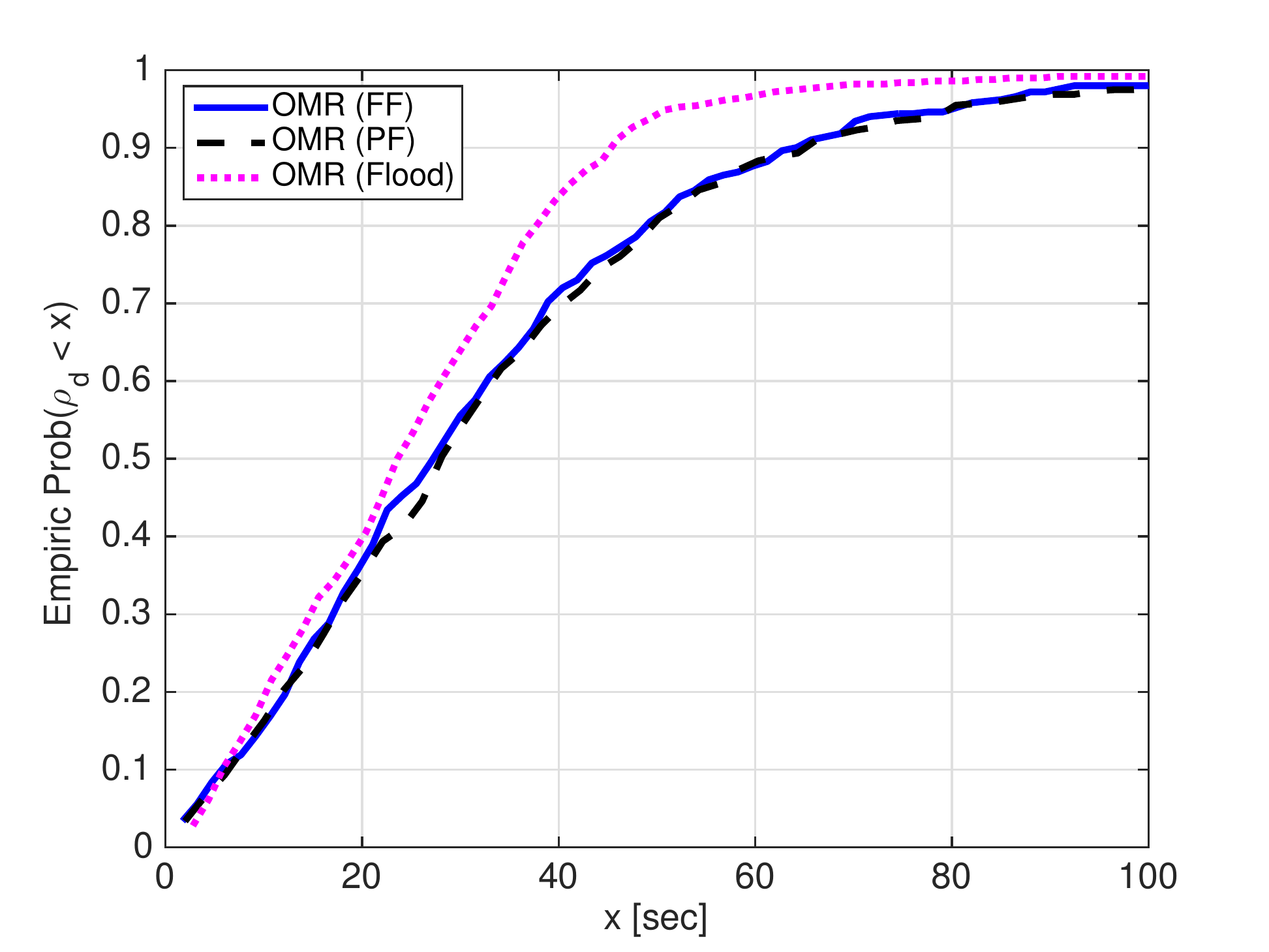}}
    \subfloat[Immediate MAC.\label{f:DelayIM}]{\includegraphics[width=0.4\textwidth]{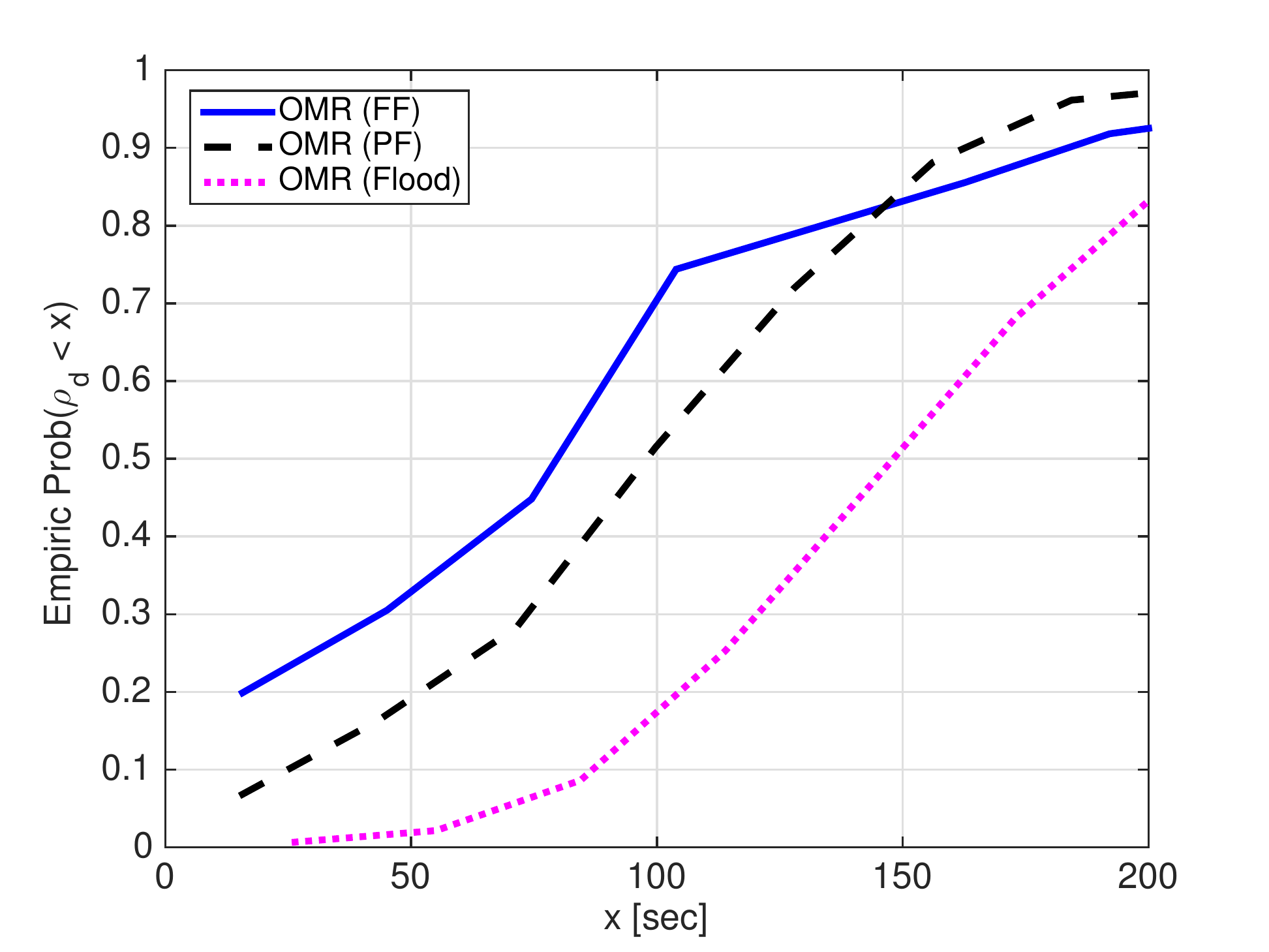}}
    \caption{Simulations: CDF of $\rho_d$ from \eqref{e:delay}. flooding achieves the best results with the Ideal MAC, which neglects collisions. The situation is reversed with the Immediate MAC, where OMR-FF achieves the best results.}\vspace{-5mm}
\end{figure}

In Figs.~\ref{f:Delay} and~\ref{f:DelayIM}, we show the cumulative distribution function (CDF) of the delay $\rho_d$ for the Ideal MAC and the Immediate (realistic) MAC, respectively. We note that the results span more than 100~s of delay. This is due to the fact that low capacity links require packets to be segmented into small fragments. This tends to increase the backlog of the nodes, which in turn increases the delay. From Fig.~\ref{f:Delay}, we observe that with Ideal MAC, the delay of the flooding scheme is better than that of OMR--FF and OMR--PF, with an average of 13.3~s compared to 16.3~s and 17~s, respectively. This result is obtained because in Ideal MAC we neglect interference, hence the flooding scheme propagates messages very fast through the network. However, when the realistic Immediate MAC is used (Fig.~\ref{f:DelayIM}), flooding's delay greatly deteriorates with respect to OMR's. We also observe that the delay of OMR-FF is better than OMR-PF's. This is because the availability of topology information makes it possible for the nodes to optimally allocate transmission resources. However, we remark that the delay of the PF version is almost as good as the FF version, while avoiding the topology information requirement. 

\begin{figure}[t]
    \centering
    \subfloat[Ideal MAC.\label{f:Goodput}]{\includegraphics[width=0.4\textwidth]{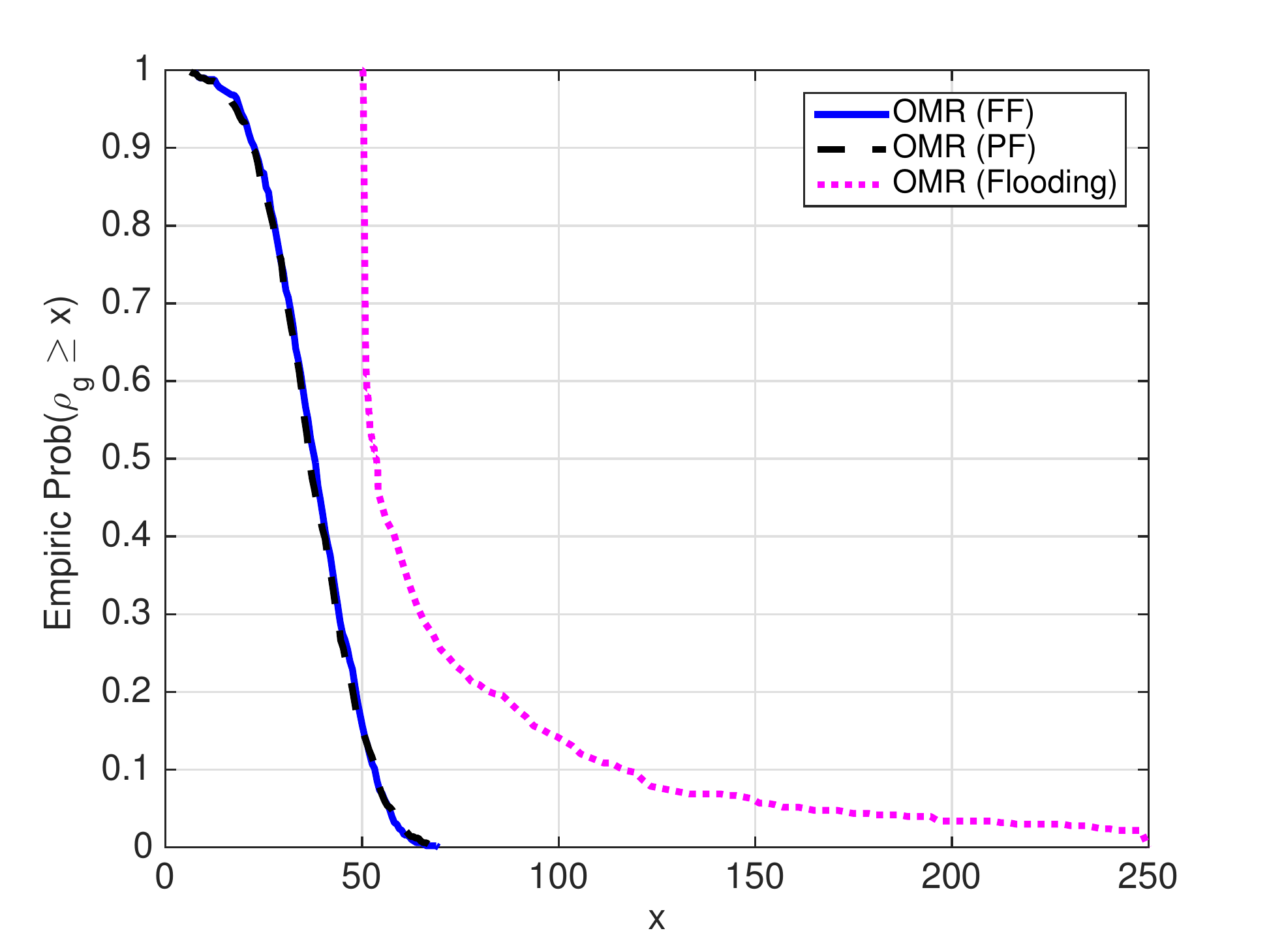}}
    \subfloat[Immediate MAC.\label{f:GoodputIM}]{\includegraphics[width=0.4\textwidth]{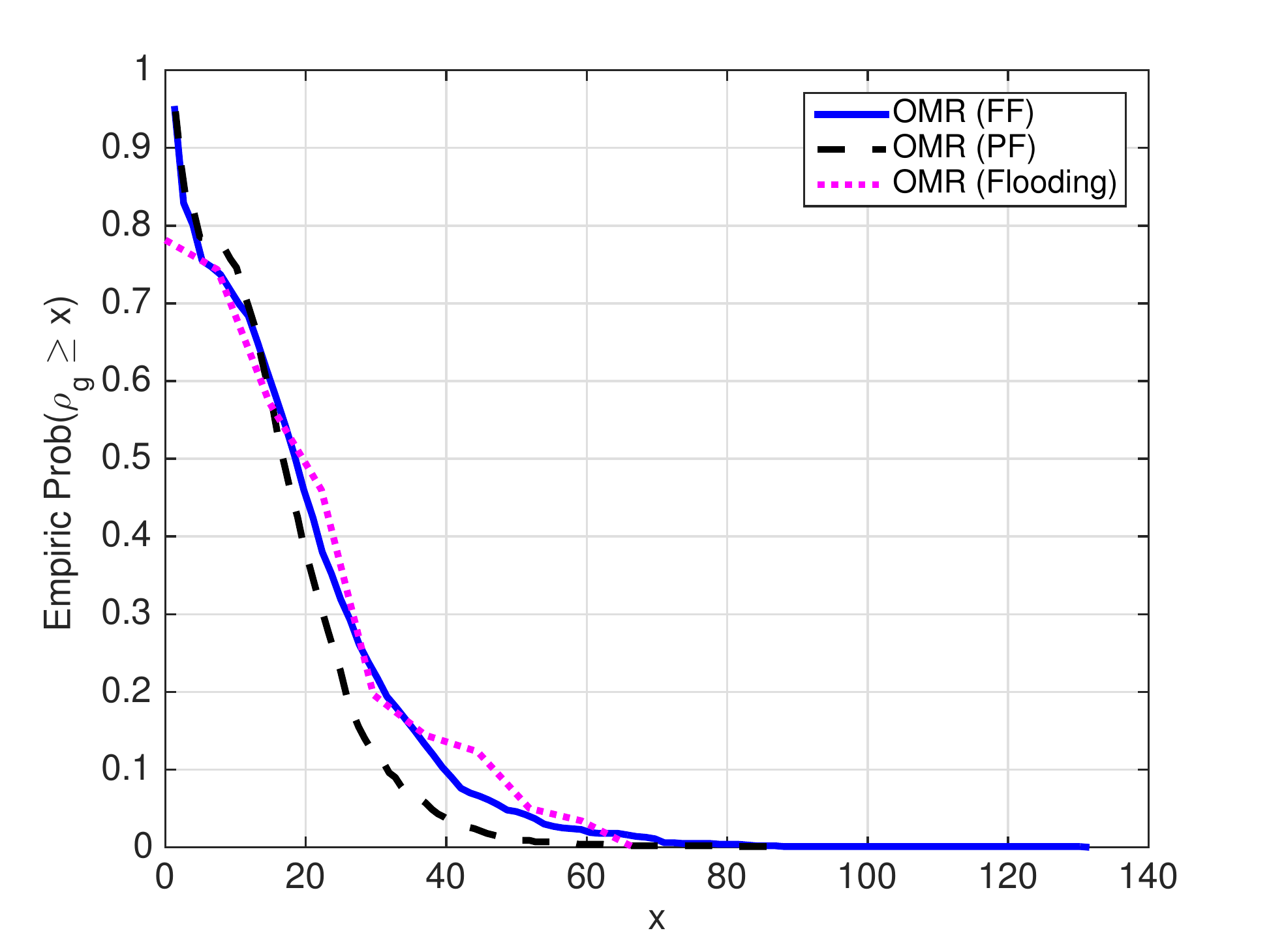}}
    \caption{Simulations: C-CDF of $\rho_g$ from \eqref{e:goodput}. Results show that due to its more transmissions OMR--PF achieves similar results to OMR--FF, and that when collisions are considered, the goodput of OMR and flooding is similar.}\vspace{-5mm}
\end{figure}

Next, in Figs.~\ref{f:Goodput} and~\ref{f:GoodputIM}, we show the complementary CDF (C-CDF) of the goodput, $\rho_g$ for the Ideal MAC and the Immediate MAC, respectively. We observe that OMR achieves almost the same performance with either MAC, implying a good level of robustness. Fig.~\ref{f:Goodput} also shows that flooding's goodput is more dispersed (meaning that OMR's performance is more predictable). The results confirm that flooding outperforms OMR only when packet collisions are ignored (Fig.~\ref{f:Goodput}), otherwise it achieves results similar to OMR (Fig.~\ref{f:GoodputIM}). As we will confirm in the lake experiment, the reason is the large number of packet collisions caused by the many transmissions of flooding.

\begin{figure}[t]
       \centering
       \includegraphics[width=0.4\textwidth]{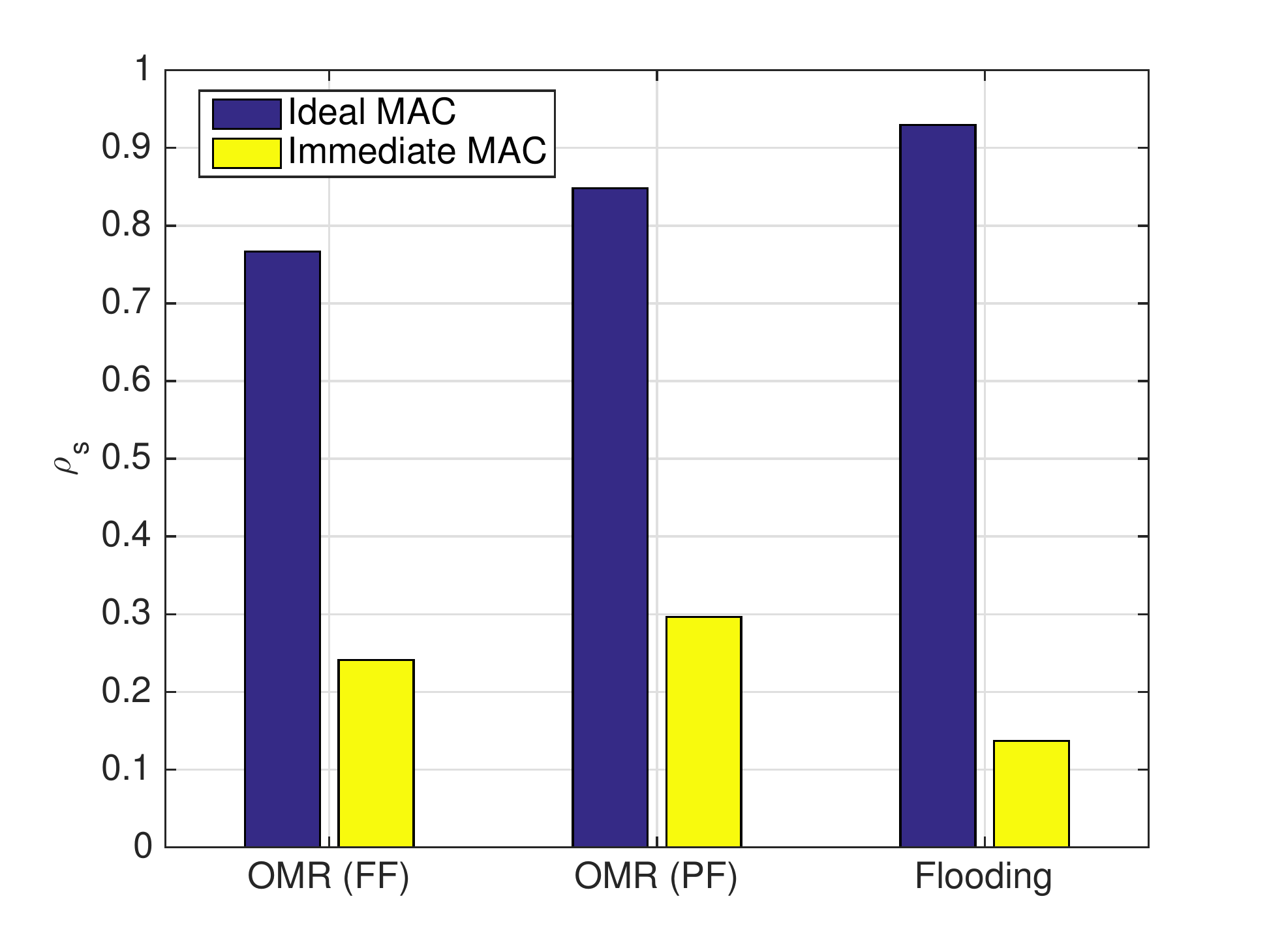}\vspace{-5mm}
       \caption{Simulations: average results of $\rho_s$ from \eqref{e:MER}. Results show that, with packet collisions, OMR--PF achieves the best results.}%
       \vspace{-5mm}
       \label{f:rxJoint}
 \end{figure}

\begin{figure}[t]
    \centering
    \subfloat[Low frequency.\label{f:UtilizationLow}]{\includegraphics[width=0.32\textwidth]{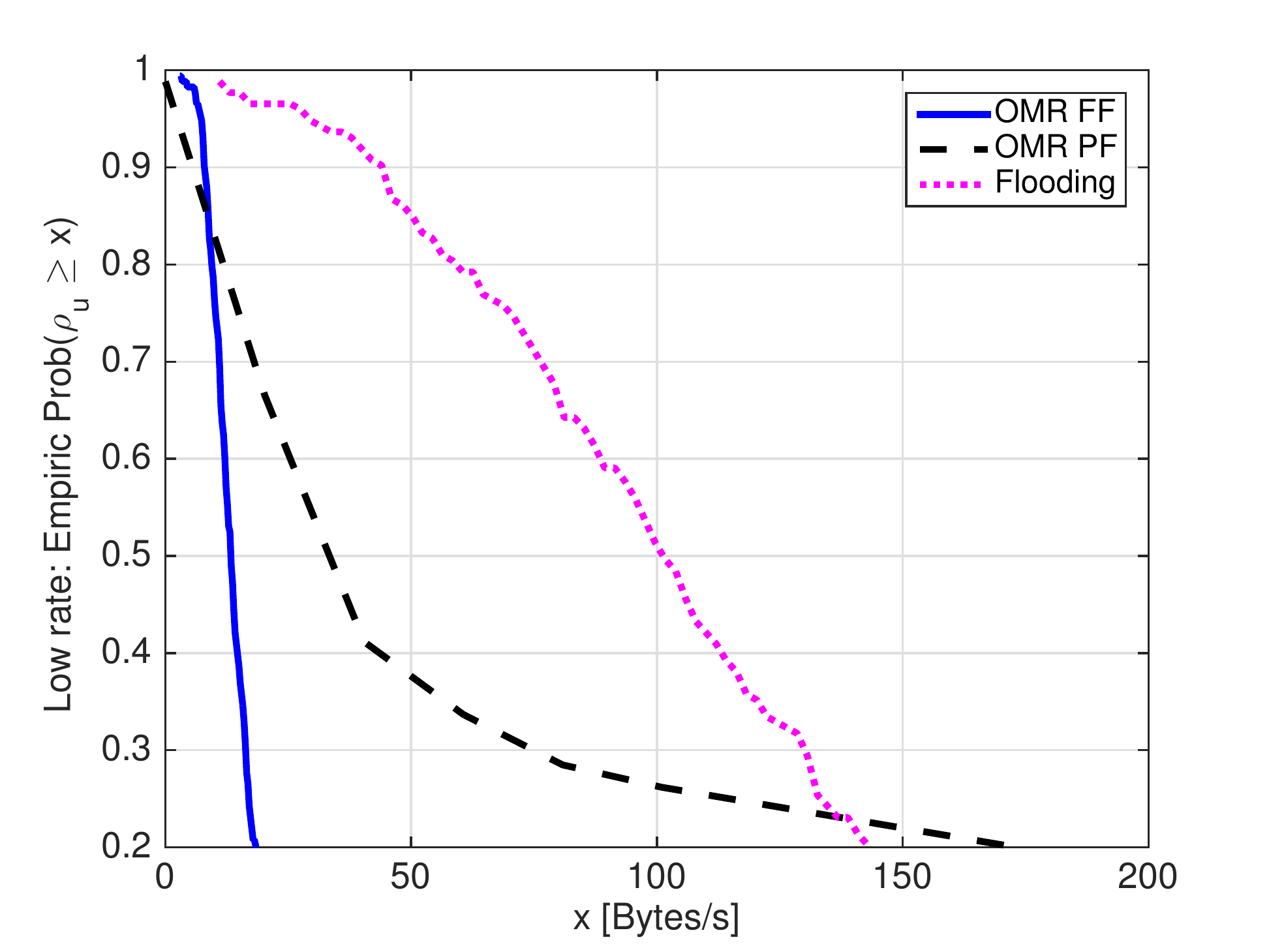}}
    \subfloat[Mid frequency.\label{f:UtilizationMid}]{\includegraphics[width=0.32\textwidth]{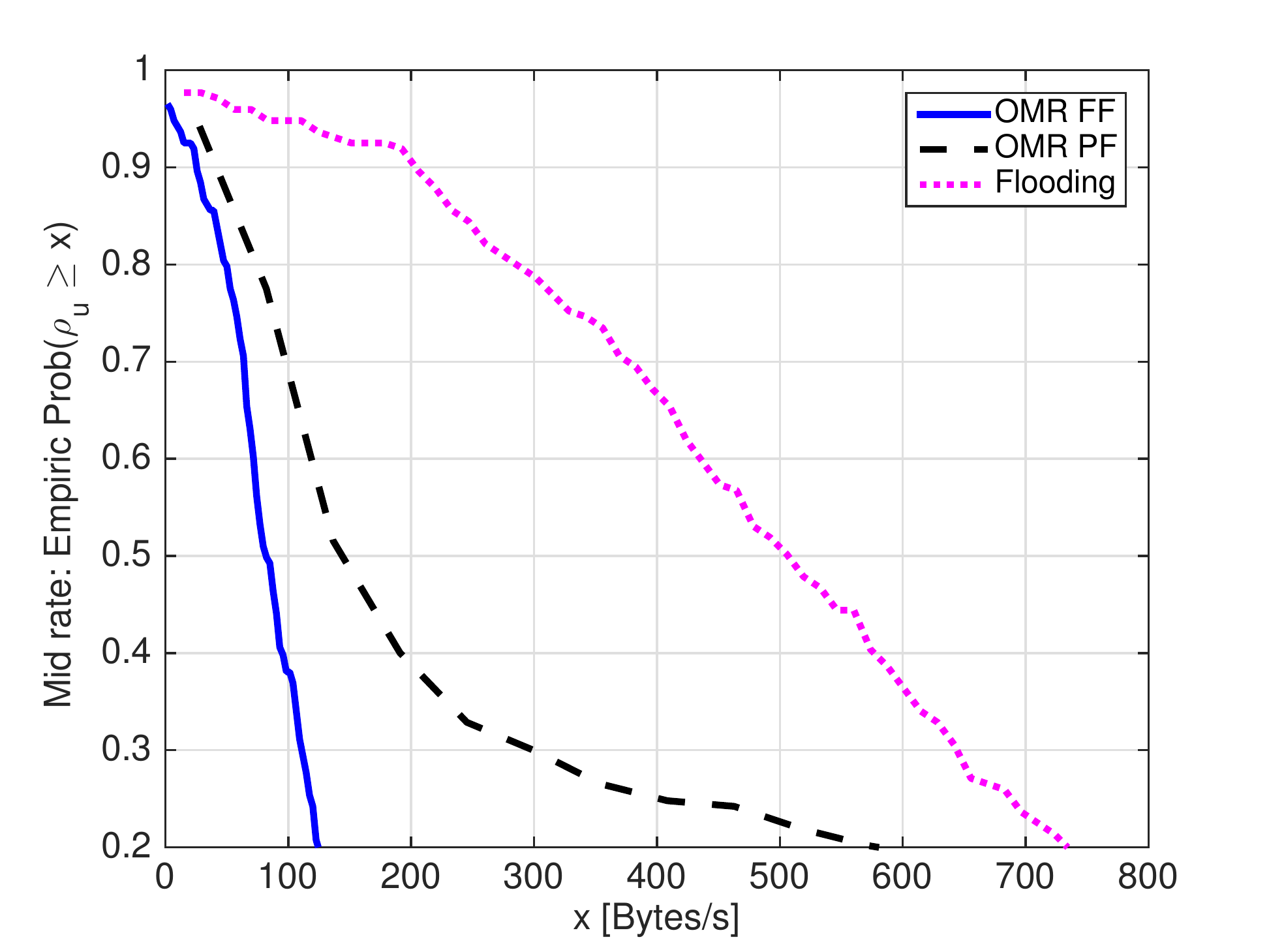}}
    \subfloat[High frequency.\label{f:UtilizationHigh}]{\includegraphics[width=0.32\textwidth]{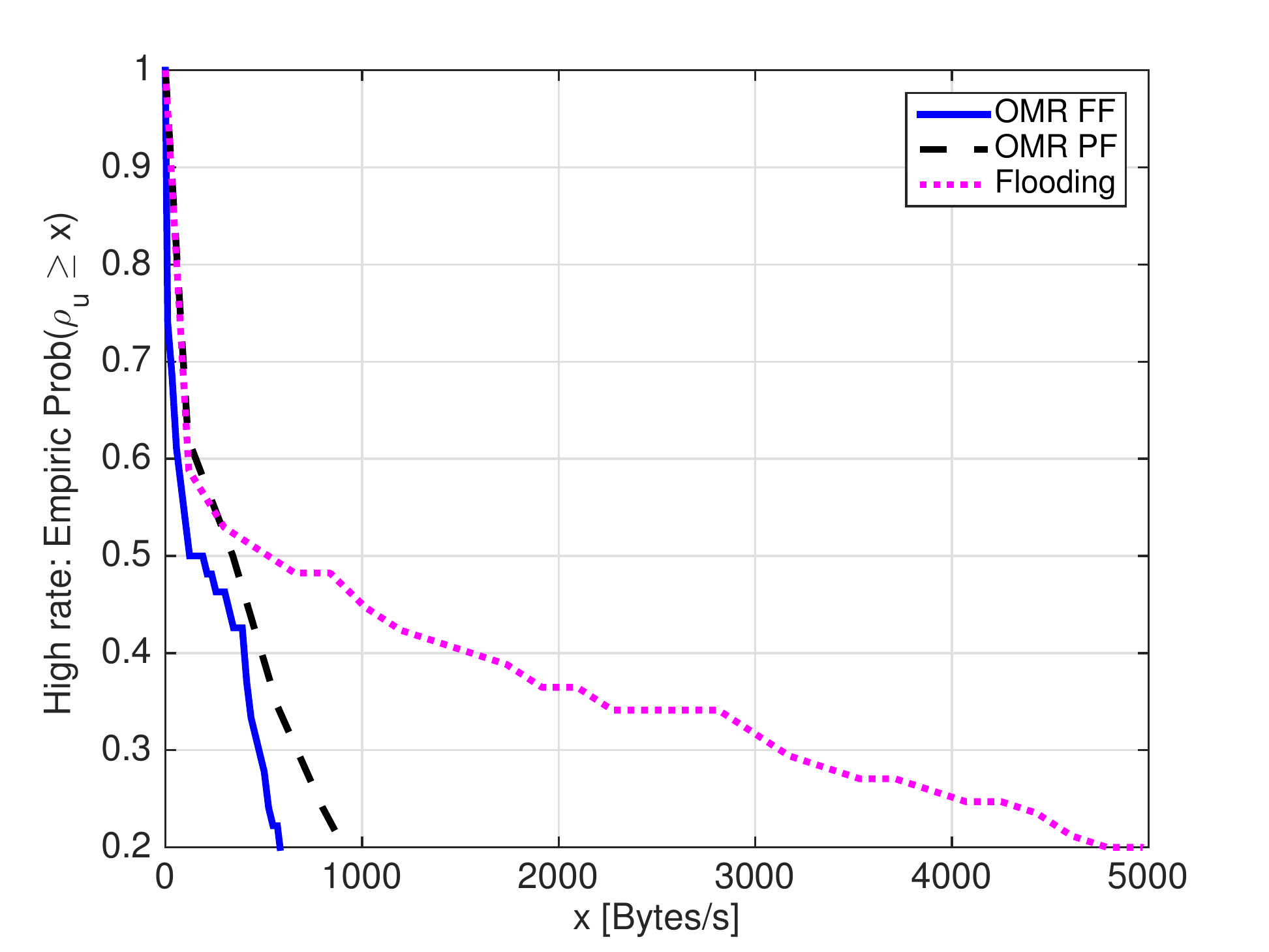}}
    \caption{Simulations: C-CDF of $\rho_u$ from \eqref{e:LU} for the Immediate MAC protocol. Without considering packet collisions, results show that flooding employs links more intensively, but its performance is highly topology dependent. OMR's behavior, on the contrary, is more consistent through different topologies.}\vspace{-4mm}
\end{figure}

When comparing the goodput of the three schemes, it is also of interest to examine the packet delivery ratio and the network fairness. The former, $\rho_s$ from \eqref{e:MER}, is shown in Fig.~\ref{f:rxJoint} for the Ideal and Immediate MACs respectively. The network fairness is measured by means of the C-CDF of the link throughput, $\rho_u$, for low rate, mid rate, and high rate transmissions (Figs.~\ref{f:UtilizationLow}, \ref{f:UtilizationMid}, and~\ref{f:UtilizationHigh}, respectively) for the case of the realistic Immediate MAC.
When the Ideal MAC is considered, Fig.~\ref{f:rxJoint} shows similar results for the three methods. However, we observe that collisions reduce flooding's delivery ratio considerably, despite the high transmission redundancy of the scheme. In fact, the increased load imposed by this redundancy on the queues of the nodes actually contributes to the poor delivery ratio of flooding. 
From Figs.~\ref{f:UtilizationLow}--\ref{f:UtilizationHigh}, we observe that since a node transmits simultaneously over all available links when using flooding, link throughput is higher in this case than with OMR. We also note that, interestingly, OMR-PF outperforms OMR-FF both in terms of the delivery ratio and in terms of link throughput. This is because the OMR-PF scheme transmits packets through more links compared to OMR-FF, and thus link throughput increases. In turn, OMR-FF which relies on full topology information, is more sensitive to link variations than OMR-PF. Hence, $\rho_s$ of OMR-PF is higher than that of OMR-FF.

\begin{figure}[t]
    \centering
    \subfloat[Average overhead ratio $\rho_o$ from \eqref{e:overhead}.\label{f:Overhead}]{\includegraphics[width=0.4\textwidth]{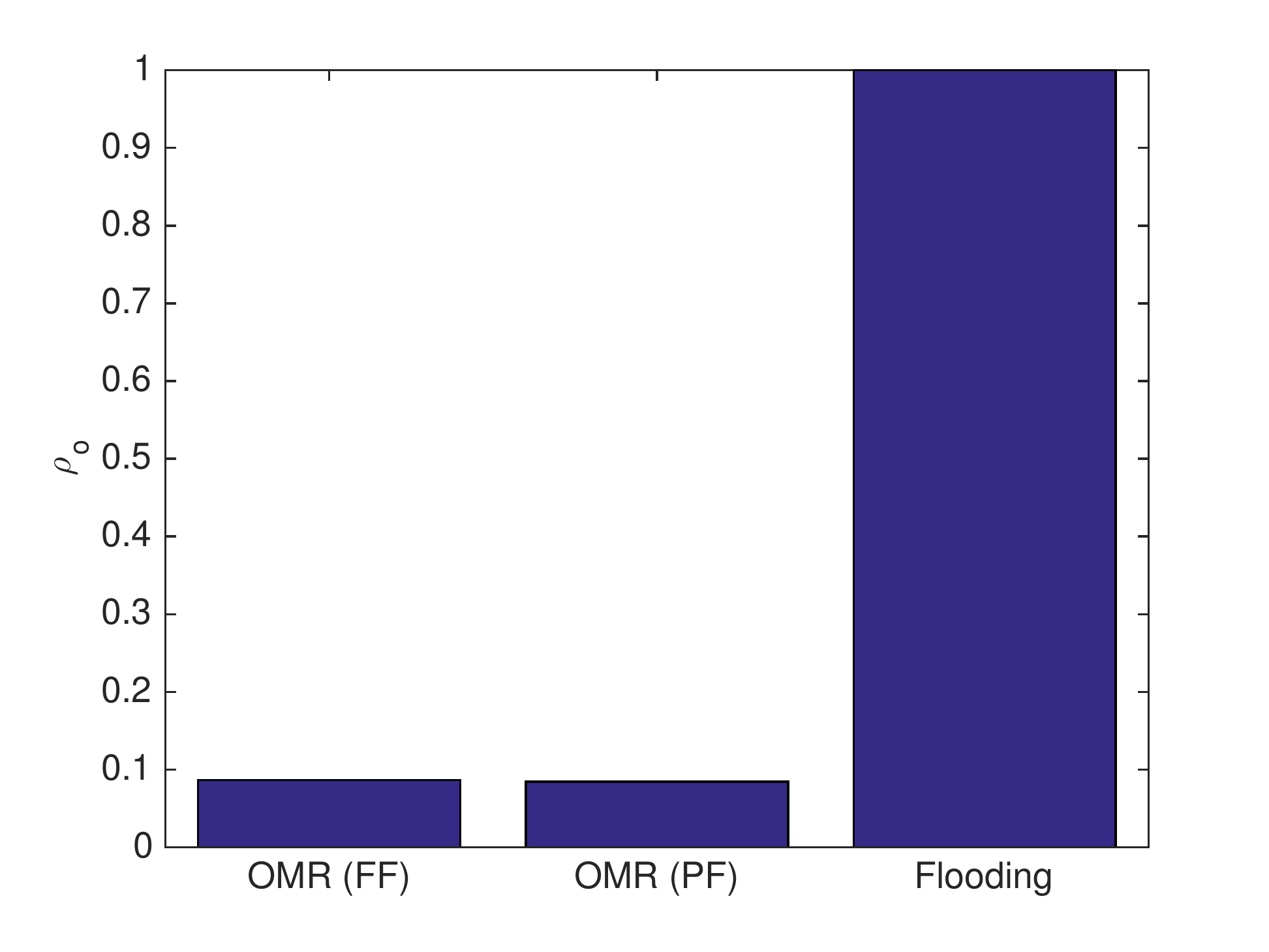}}
    \subfloat[Average $\rho_e$ from \eqref{e:tx}.\label{f:Total}]{\includegraphics[width=0.4\textwidth]{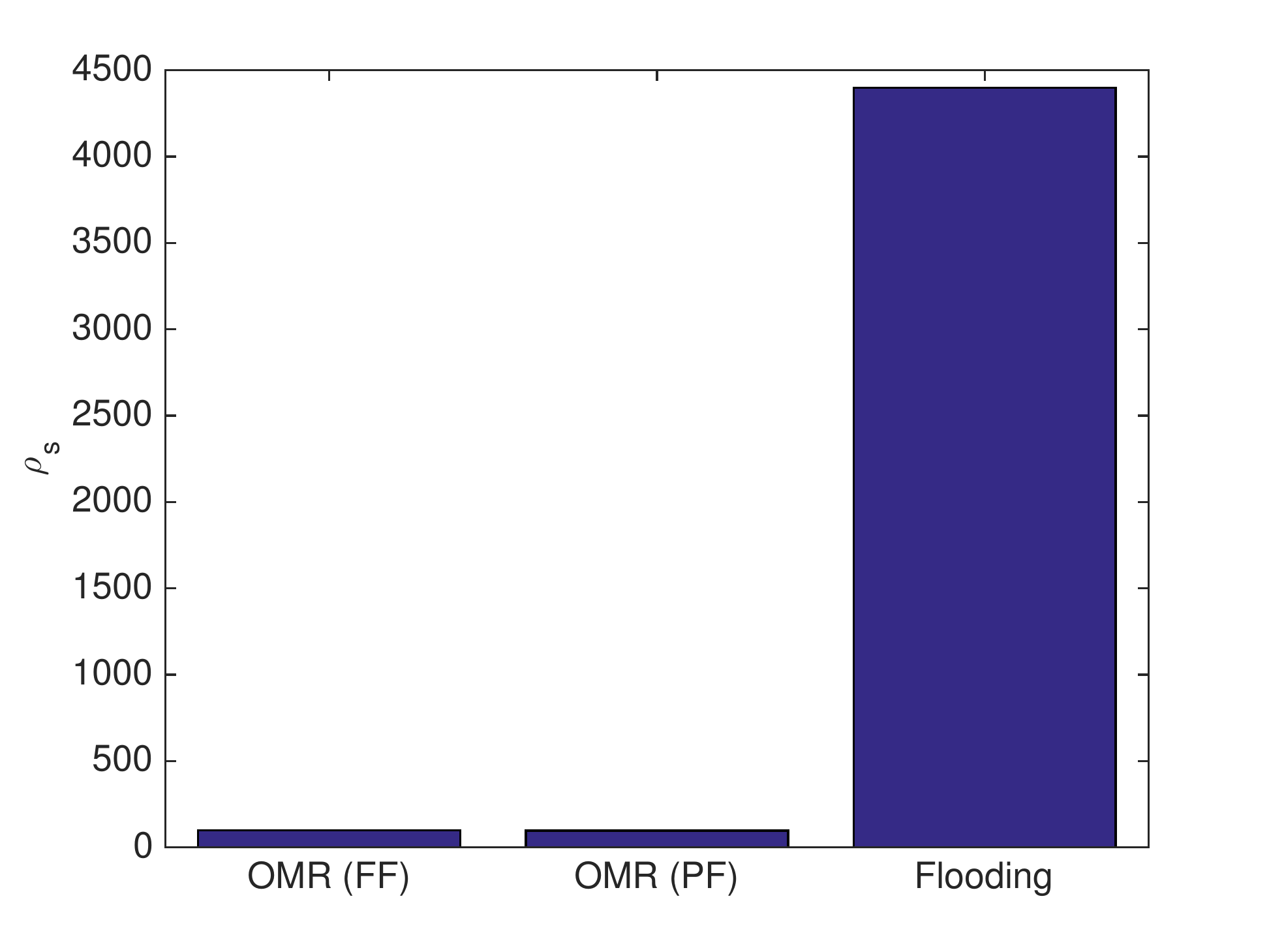}}
    \caption{Simulations. Energy efficiency for the Immediate MAC protocol. OMR is more efficient than flooding. OMR-FF and -PF perform similarly.}\vspace{-5mm}
\end{figure}

To comment on the energy efficiency of the three methods, in Figs.~\ref{f:Overhead} and~\ref{f:Total} we show 
overhead $\rho_o$, and the total number of transmitted bytes, $\rho_e$, respectively, for the case of Immediate MAC. While multiple (redundant) copies of all messages are received with the flooding scheme, in the two OMR versions the sink receives extra copies only for about 8$\%$ of the messages. 
This result is further emphasized by the huge difference in the total number of bytes sent as shown in Fig.~\ref{f:Total}. Comparing the goodput performance from Fig.~\ref{f:Goodput}, the overhead results, and the total number of Bytes transmitted of OMR-FF and of OMR-PF, we observe almost identical performance, and that only a few redundant transmissions exist. Also here, this result works in slight favor of OMR-PF which requires much less knowledge and is thus more distributed.

%% file: experiment.tex
\section{Sea experiment} \label{sec:trial}

Our simulations revealed that the comparison between flooding and OMR depends on packet collisions. 
To complement these conclusions, results from a real underwater environment are needed.
Experimenting with real systems includes non-ideal modem hardware behaviors, multipath propagation, actual packet collisions, the impact of finite memory in each node, and delays due to the management of multi-modal technologies. In the following, we describe the setup of our field trial and the results obtained. 

\subsection{Setup of the experiment} \label{sec:trial.setup}

The trial took place in June 2016, in the Werbellin lake, north of Berlin, Germany. The lake is narrow and long, with a maximum depth of 55~m. This environment produces a channel with a long delay spread and location dependent ambient noise, which poses a significant challenge for underwater acoustic networks. The experiment included six nodes which were deployed at four different geographical locations. Three locations were reached using small vessels: two motorized inflatable boats, and one motorboat. The fourth location was one of the lake's docks. Throughout the experiment, the boats tended to drift at an approximate speed of 0.25~m/s.
 
The multi-modal functionality was obtained via three types of EvoLogics underwater acoustic modems~\cite{evologics_modems_page}. A low-rate, low-frequency technology was incorporated by the S2C 18-34 modem, 
\begin{wrapfigure}[10]{l}[0pt]{2.5in}
    \centering
    \vspace{-3mm}
    \includegraphics[width=2.5in]{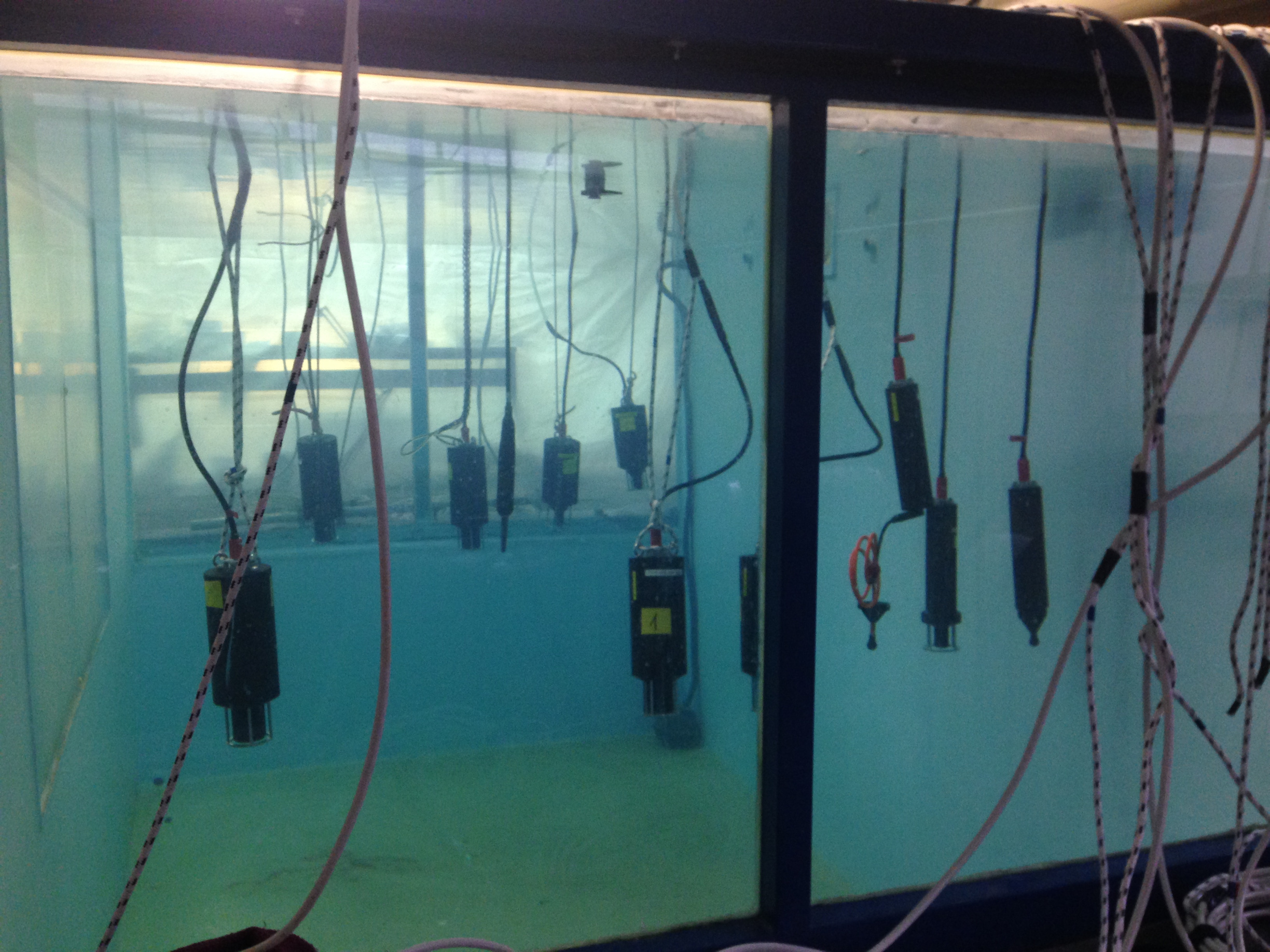}
    \caption{Experiment: A picture taken in a water tank showing the ten underwater acoustic modems during preliminary system tests.}
    \label{fig:modems_tank}
\end{wrapfigure}
having a maximum transmission range of 3.5~km. As a shorthand, we dub this technology low-frequency (\textit{LF}) in the following. A second type of communication technology was obtained via the S2C 48-78 modem, which has a maximum range of 1~km (mid-frequency technology, \textit{MF}). The third type of communication technology was represented by the S2CM HS model, which is employed over short links of up to 300~m (high frequency technology, \textit{HF}). System pre-tests in a tank using the minimum source level allowed (see picture in Fig.~\ref{fig:modems_tank}) revealed that the modems could work in parallel without significant outband interference. In total, 
we used 10 modems: five LFs, three MFs, and two HF. 

\begin{figure}[t!]
    \centering
    \subfloat[Topology 1\label{subfig:exp.topo1}]%
        { 
          \includegraphics[width=0.416\textwidth]{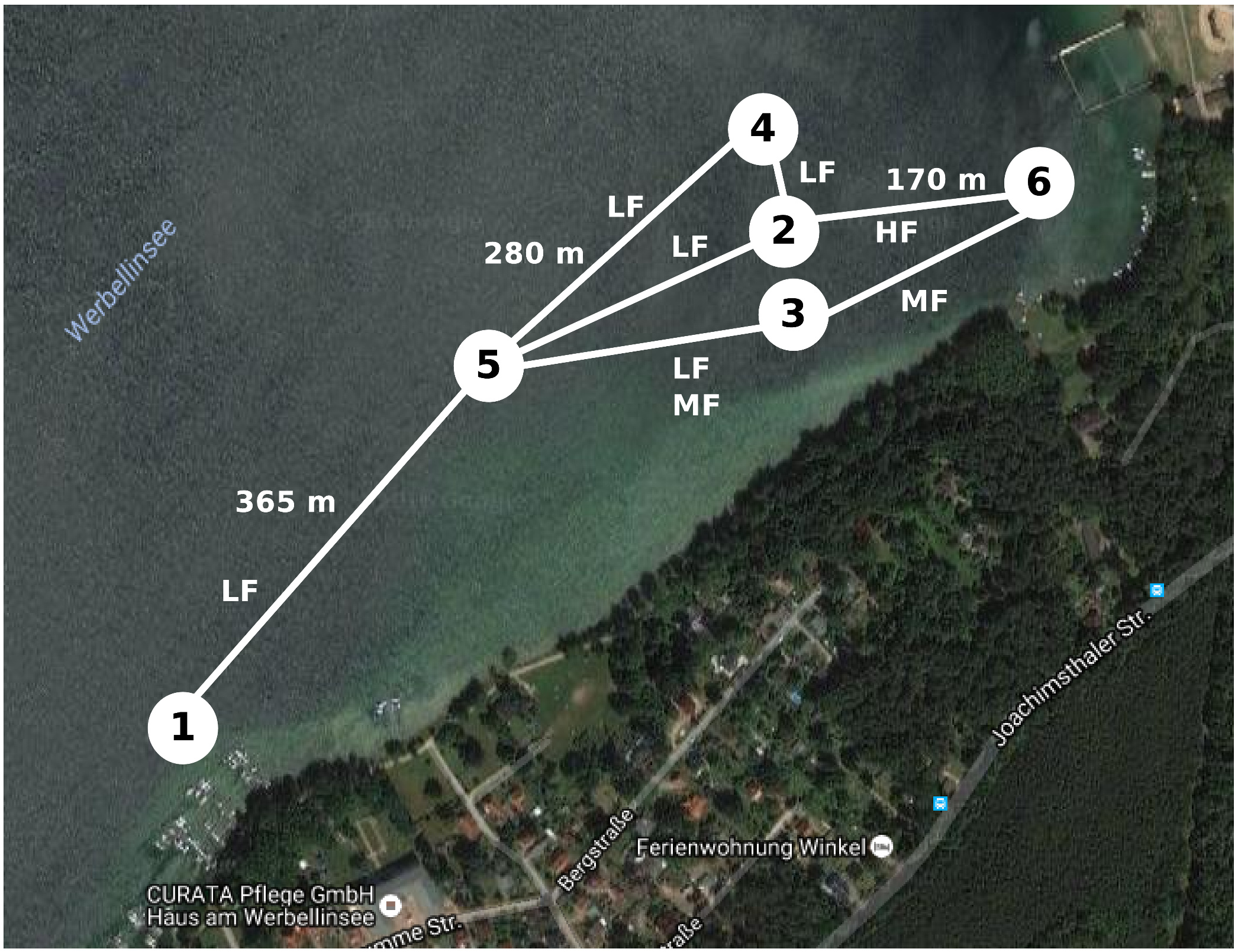}%
        }
    \hspace{3mm}
    \subfloat[Topology 2\label{subfig:exp.topo2}]%
        { 
          \includegraphics[width=0.416\textwidth]{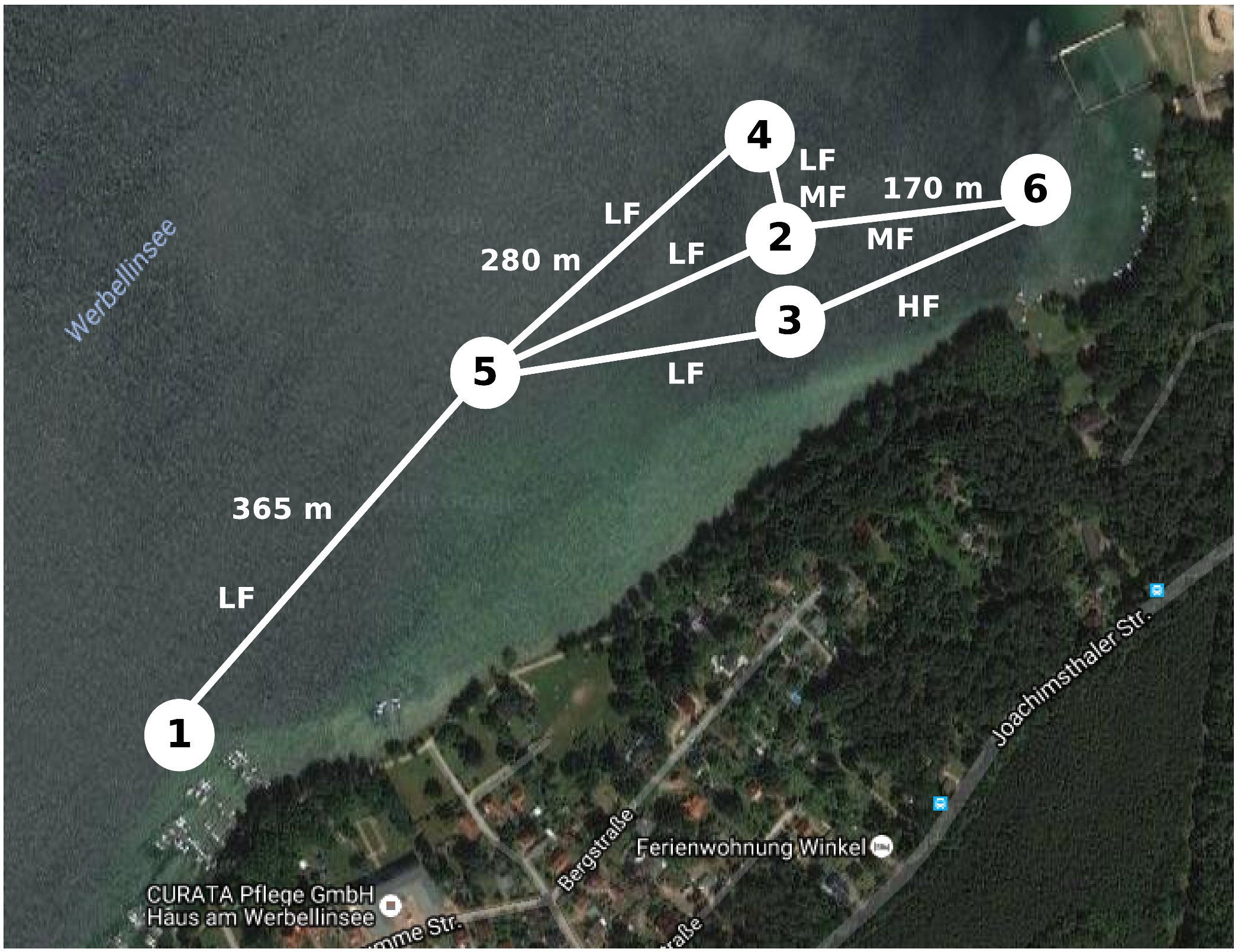}%
        }\\[2mm]
    \subfloat[Topology 3\label{subfig:exp.topo3}]%
        { 
          \includegraphics[width=0.27\textwidth]{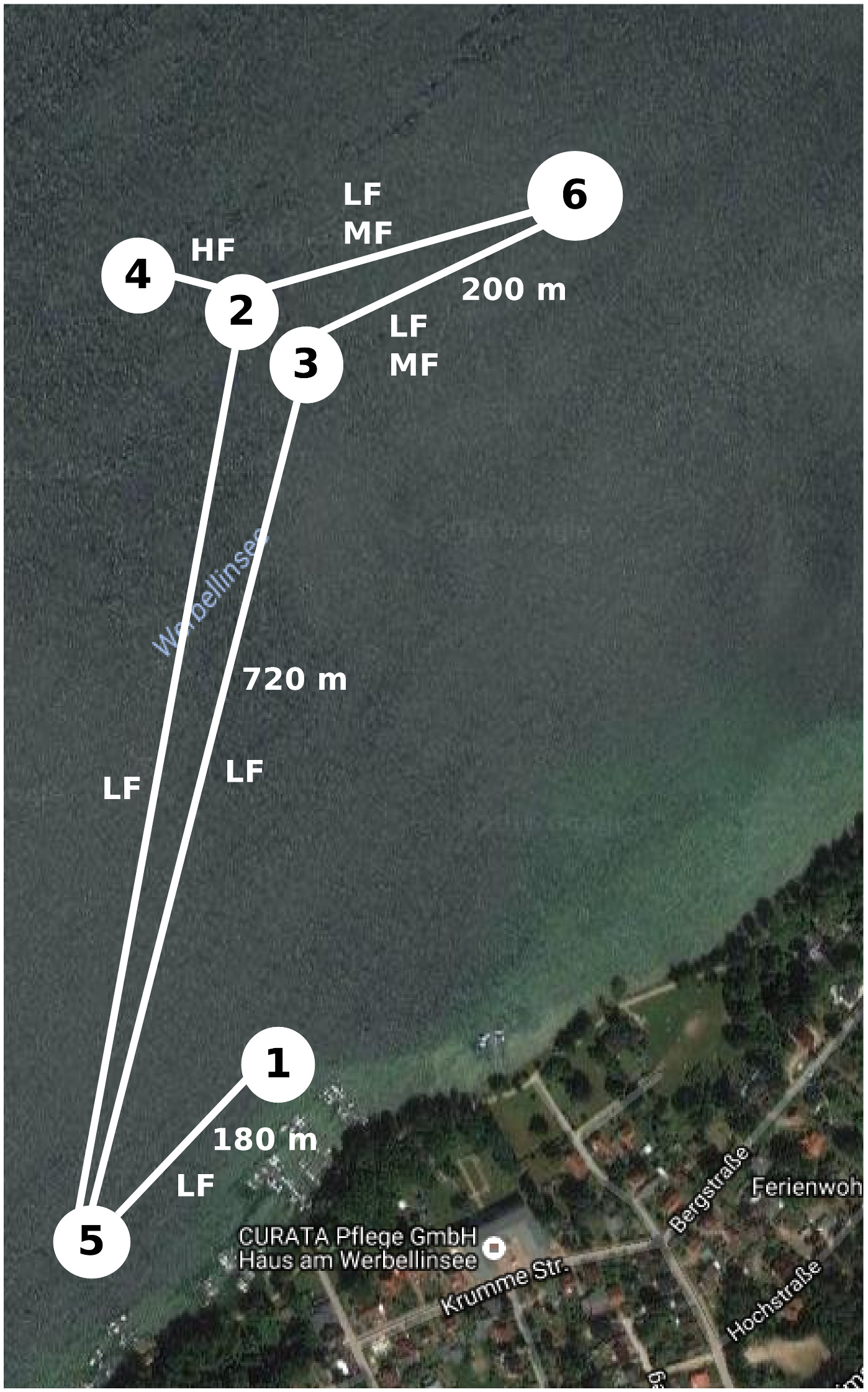}%
        }
    \hspace{2mm}
    \subfloat[Topology 4\label{subfig:exp.topo4}]%
        { 
          \includegraphics[width=0.27\textwidth]{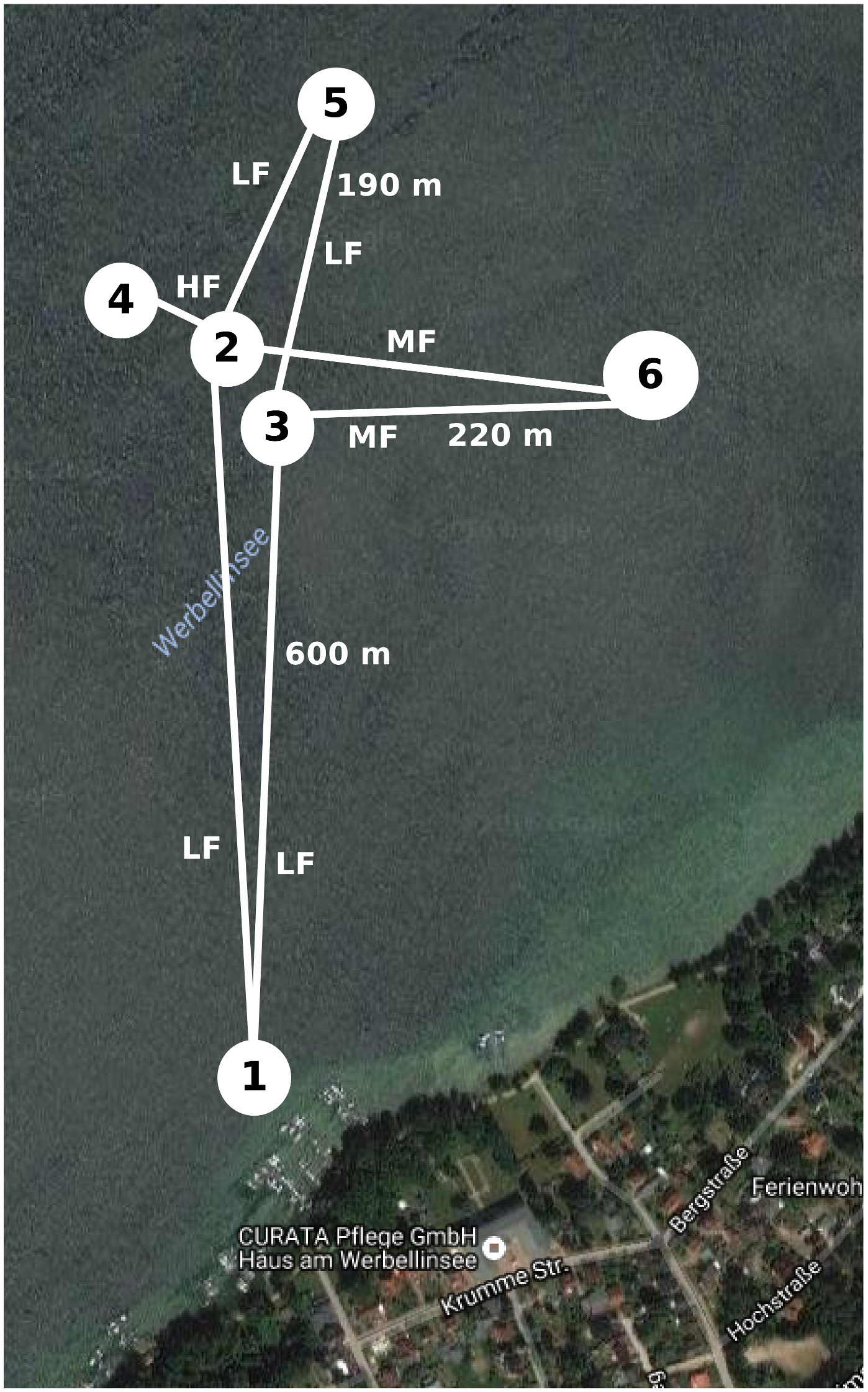}%
        }
    \hspace{2mm}
    \subfloat[Topology 5\label{subfig:exp.topo5}]%
        { 
          \includegraphics[width=0.27\textwidth]{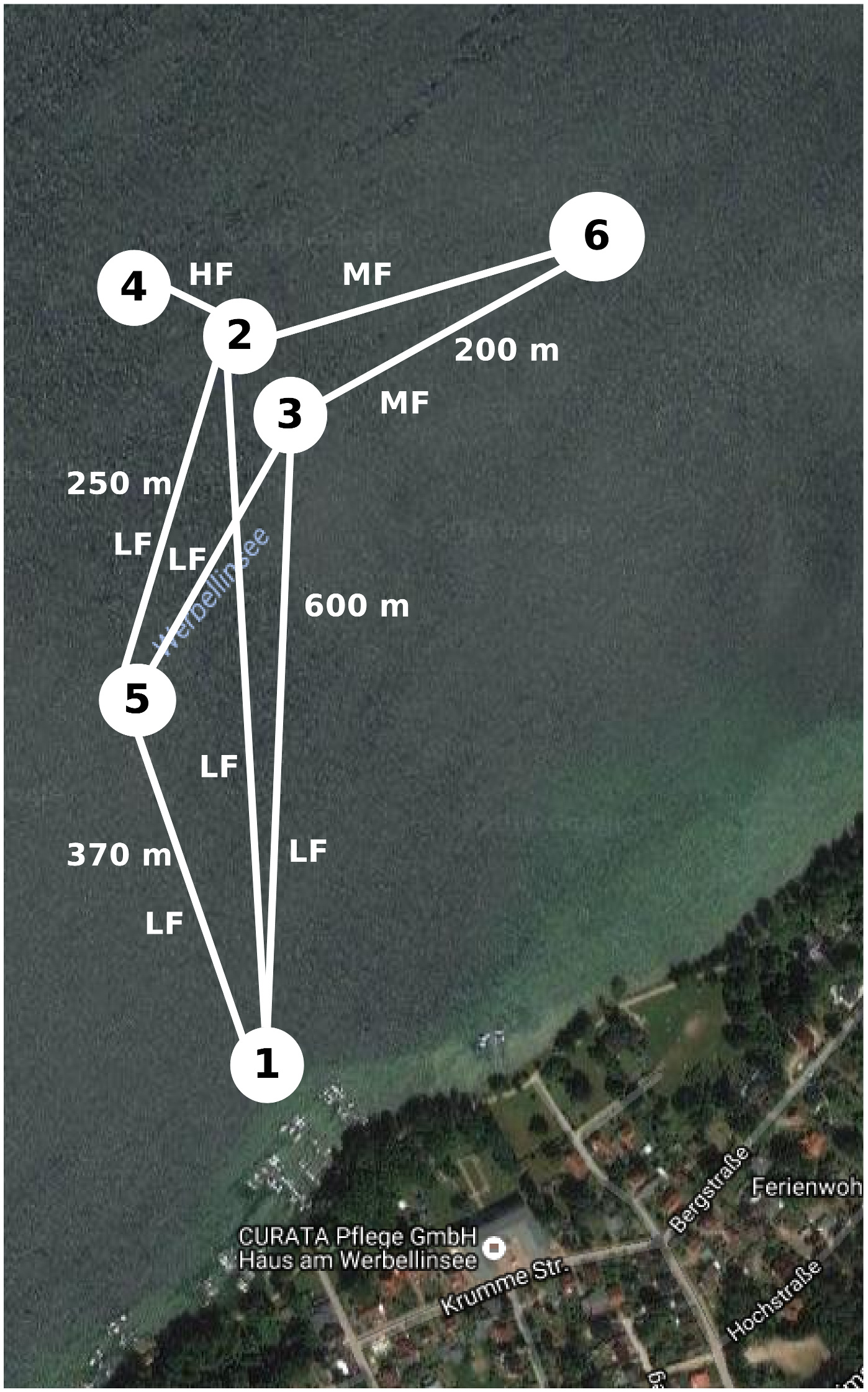}%
        }
    \caption{Logical network topology configurations and locations of the nodes in the five scenarios considered in our lake experiment. Each link is tagged with the technologies that can be used over that link.}\vspace{-5mm}
    \label{fig:exp}
\end{figure}

The performance of the routing schemes was tested in five different network topologies. In Fig.~\ref{fig:exp}, we illustrate the topologies tested, where solid lines represent a communication link, and we mark the LF, MF, and/or HF communication technologies used by each node. %
To form these topologies, the boats moved between several waypoints in the lake as shown in Fig.~\ref{fig:exp}. The figure also shows the location of each of the four stations, the distance between the stations, and the ID of the nodes\newpage
\newcommand{\pcnl}{\newline[-3mm]}
\begin{wraptable}[11]{r}[0cm]{11.2cm}%
    \centering
    \vspace{-3mm}
    \caption{Technologies available to each node in each scenario and approximate deployment depth (between parentheses)}
    \label{tab:techspernode}
    \vspace{-5mm}
    \footnotesize
    \begin{tabular}{@{\hspace{1mm}}L{\widthof{Node M}}*{5}{L{\widthof{Tolopogy M}}}@{\hspace{1mm}}}
                     & {\bf Topology 1} & {\bf Topology 2} & {\bf Topology 3} & {\bf Topology 4} & {\bf Topology 5} \\
        \midrule 
        {\bf Node 1} & MF, HF\pcnl(3~m)    & MF, HF\pcnl(3~m)    & LF, MF\pcnl(3~m)        & MF\pcnl(3~m)            & MF\pcnl(3~m)           \\
        {\bf Node 2} & LF, HF\pcnl(10~m)   & LF, MF\pcnl(10~m)   & LF, MF, HF\pcnl(10~m)   & LF, MF, HF\pcnl(10~m)   & LF, MF, HF\pcnl(10~m)  \\
        {\bf Node 3} & LF, MF\pcnl(10~m)   & LF, HF\pcnl(10~m)   & LF, MF\pcnl(10~m)       & LF, MF\pcnl(10~m)       & LF, MF\pcnl(10~m)      \\
        {\bf Node 4} & LF\pcnl(10~m)       & LF, MF\pcnl(10~m)   & HF\pcnl(10~m)           & HF\pcnl(10~m)           & HF\pcnl(10~m)          \\
        {\bf Node 5} & LF\pcnl(5~m)        & LF\pcnl(10~m)       & LF\pcnl(5~m)            & LF\pcnl(10~m)           & LF\pcnl(10~m)          \\
        {\bf Node 6} & LF\pcnl(10~m)       & LF\pcnl(10~m)       & LF\pcnl(10~m)           & LF\pcnl(10~m)           & LF\pcnl(5~m)           \\
        \bottomrule
    \end{tabular}
\end{wraptable}
\noindent in each station. As shown in the figure, node~6 was always deployed on the pier, nodes~1 and~5 were hosted in one inflatable boat each, whereas nodes~2, 3 and~4 were deployed from opposite ends of the motorboat. 
In all topologies, node~6 served as the sink node. The nodes deployed the modems at a depth of roughly one half of the local water column depth. 
%
Table~\ref{tab:techspernode} shows the mapping between the nodes and the available technologies in each scenario, along with the approximate deployment depth. Most of the required reconfigurations involve the shorter-range technologies MF and HF available to nodes~2, 3 and~4.
Each node was driven by a laptop which ran the routing logic and drove the modems. This was achieved by a novel combination of Matlab and DESERT~\cite{ucomms16_desert}, adapted to manage multi-modal technologies. Note that routing was performed in a distributed fashion.

For each of the five topologies, we conducted three 10-min experiments, one employing OMR--FF, one with OMR--PF, and one with flooding. The information regarding the communication technologies available in each topology (one-hop links for OMR--PF, or full topology information for OMR--FF) was obtained via a preliminary link discovery phase~\cite{ted_roee_2016}%
	\footnote{Note that the link discovery is not regarded as an overhead for the routing scheme, as we assume an underlying MAC protocol that handles both link discovery and transmissions}.   %

Each node generated its own set of data packets, which remained equal throughout the experiments. This traffic was generated according to a Poisson process of rate $\lambda=2$ packets per minute per node. The size of each packet was set uniformly at random between~0 and 64~kbit. During each experiment, the nodes sent the data packets through multiple hops towards the sink, abiding to the rules of the OMR protocol presented in Section~\ref{sec:routing} or the flooding scheme mentioned in Section~\ref{sec:sim}. 
Periodically, the nodes exchanged information related to the number of packets in their queue, their neighbor lists and the remaining information needed to run the protocol. When operating OMR, the reception of each data packet was separately acknowledged. In case an acknowledgment was not received, the packet was re-transmitted up to two times by the modem's MAC protocol. Broadcast packets (e.g., reporting the queue status in the OMR protocol and the hop history in the flooding protocol) were not acknowledged.

\subsection{Results}\label{sec:trial.res}

With five topologies tested, we measure the performance of the experiment in terms of the end-to-end transmission delay $\rho_d$ in \eqref{e:delay}, the goodput $\rho_g$ in \eqref{e:goodput}, and the link throughput $\rho_u$ in \eqref{e:LU}. The end-to-end delay of each message was calculated only once the sink (node~1) received the message in full, while the goodput was calculated for each message segment received by the sink. For the link throughput, we considered any successful transmission in the link regardless of whether the packet segment was ultimately received by the sink or not. 

\begin{figure}
    \centering
    \subfloat[Flooding]{\includegraphics[width=0.225\textwidth]{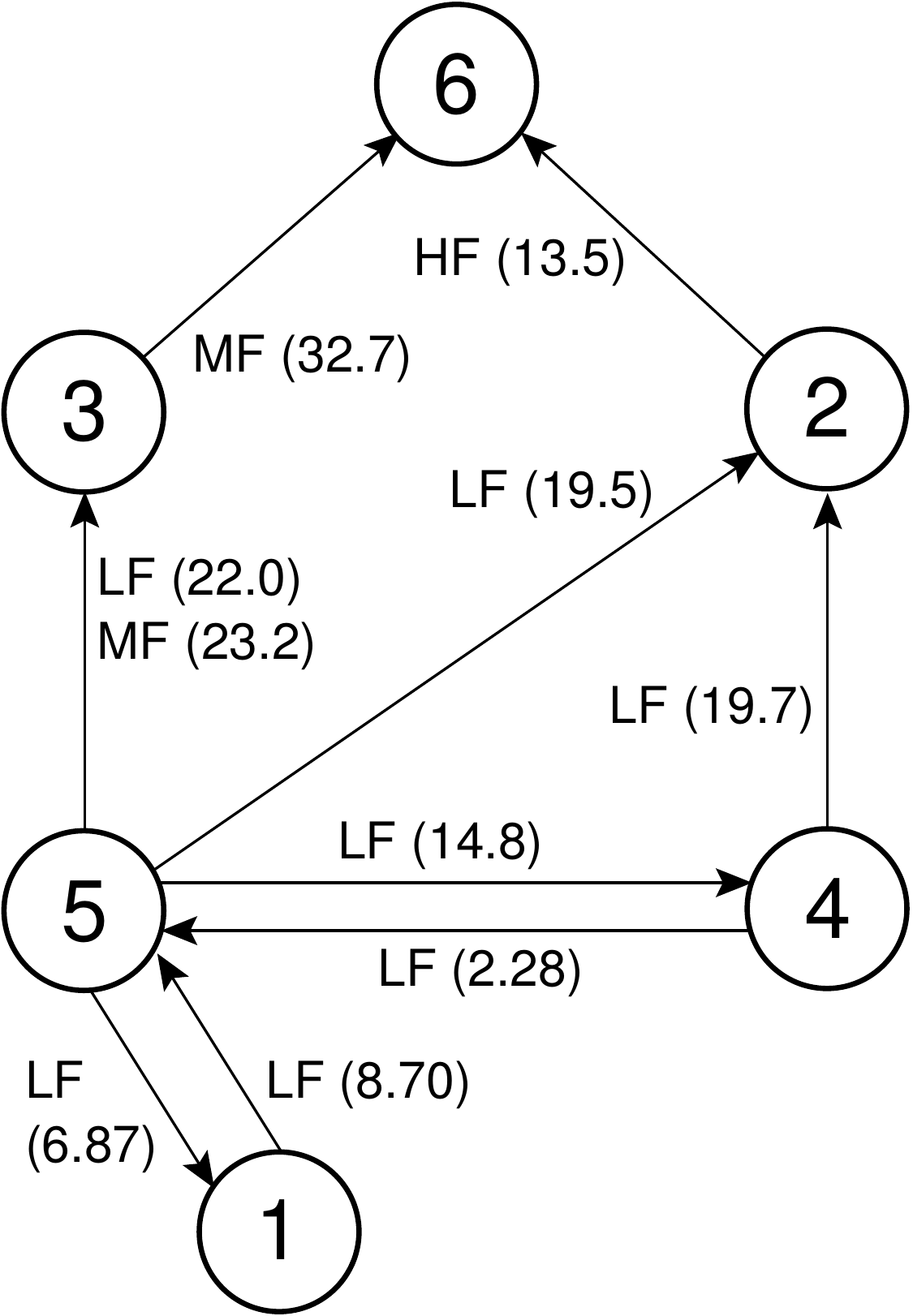}   \label{fig:rxbits.flood}}
    \hspace{7.5mm}
    \subfloat[OMR--PF]{\includegraphics[width=0.225\textwidth]{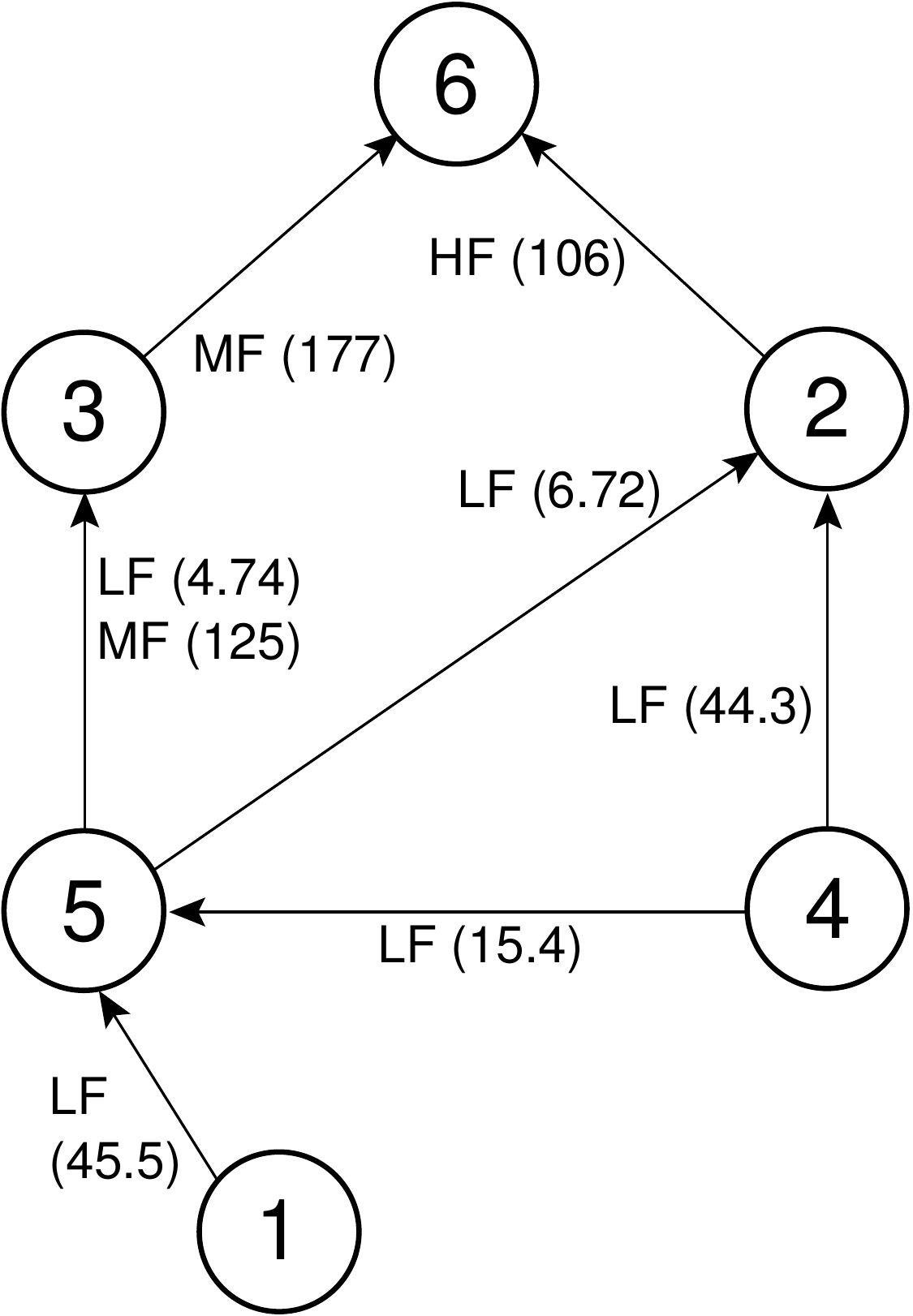}   \label{fig:rxbits.notopo}}
    \hspace{7.5mm}
    \subfloat[OMR--FF]{\includegraphics[width=0.225\textwidth]{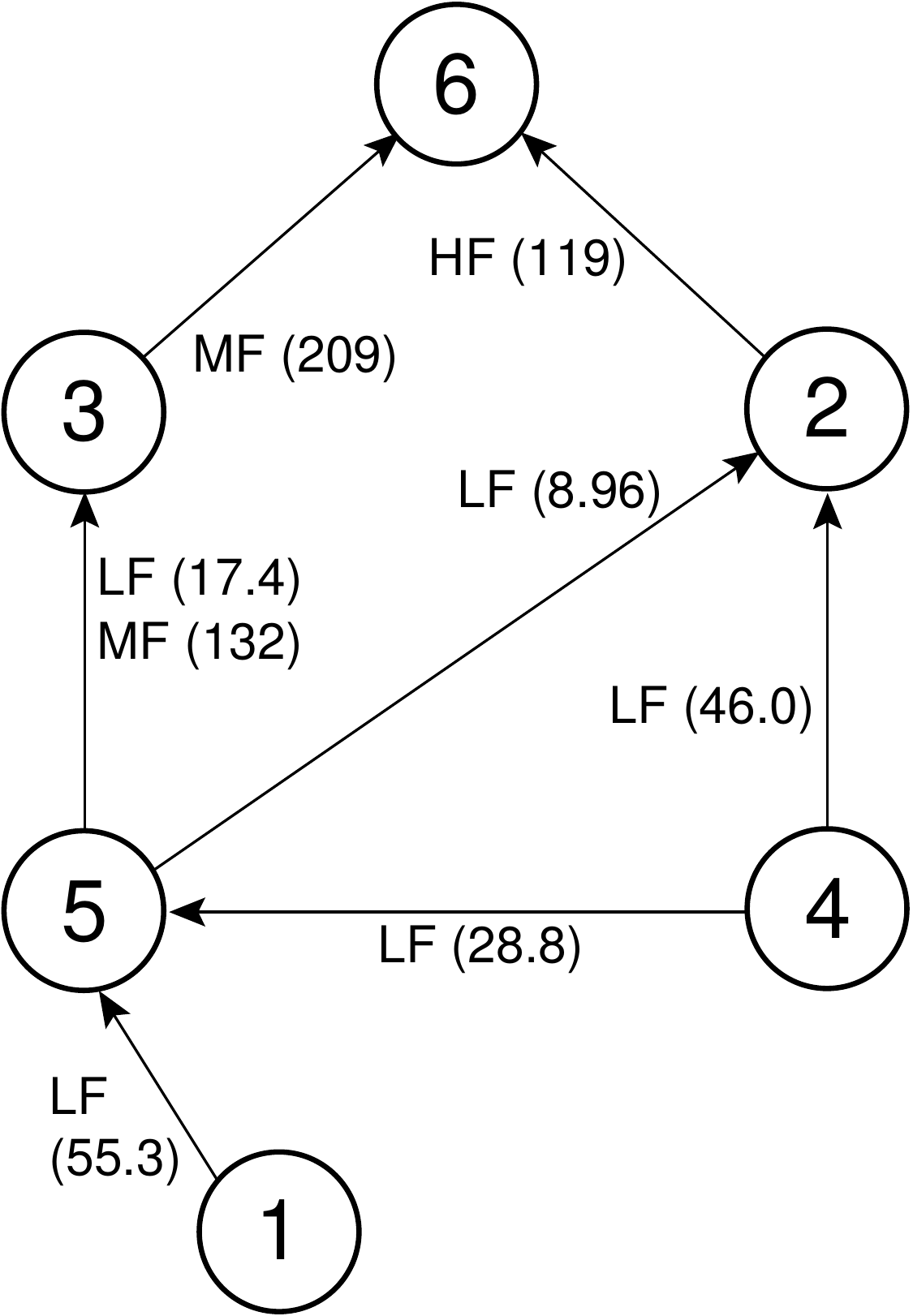}   \label{fig:rxbits.withtopo}}
    \caption{Link throughput in kbit/s for all protocols run in the experiment, Topology~1.}\vspace{-5mm}
    \label{fig:rxbits}
\end{figure}

We initially focus on Topology~1 (see Fig.~\ref{fig:exp}), and start by discussing a link throughput sample in Fig.~\ref{fig:rxbits}. We observe that flooding is too aggressive in transmitting packets over all available links, and results in poor link throughput (Fig~\ref{fig:rxbits.flood}). 
By crossing transmission and reception logs we note that the main reason is that flooding is subject to a high chance of collisions, and to the high bit error rate that results. On the contrary, the two OMR versions convey traffic more reliably through the network, resulting in higher throughput. In particular, OMR--PF (that has no access to topology information beyond first-hop neighbors) tends to be more conservative (Fig~\ref{fig:rxbits.notopo}). As a consequence, the throughput of node~3's MF link and of the LF links of nodes~1, 4 and~5 are limited. Full topology awareness in OMR--FF makes nodes~1 and~4 aware of the capacity of node~5's upstream links, so that they can push more traffic through their LF links to node~5. In turn, node~5 will convey this to 3 through both the MF and the LF links, and finally to node~6 through node~3's MF link. The overall result is better link throughput over all technologies (Fig~\ref{fig:rxbits.withtopo}) and a larger number of packets that reach the sink (node~6) correctly.

\begin{figure}[t]
    \centering
    \subfloat[$\rho_d$ (see Eq.~\eqref{e:delay})\label{f:ExpDelay}]{\includegraphics[width=0.4\textwidth]{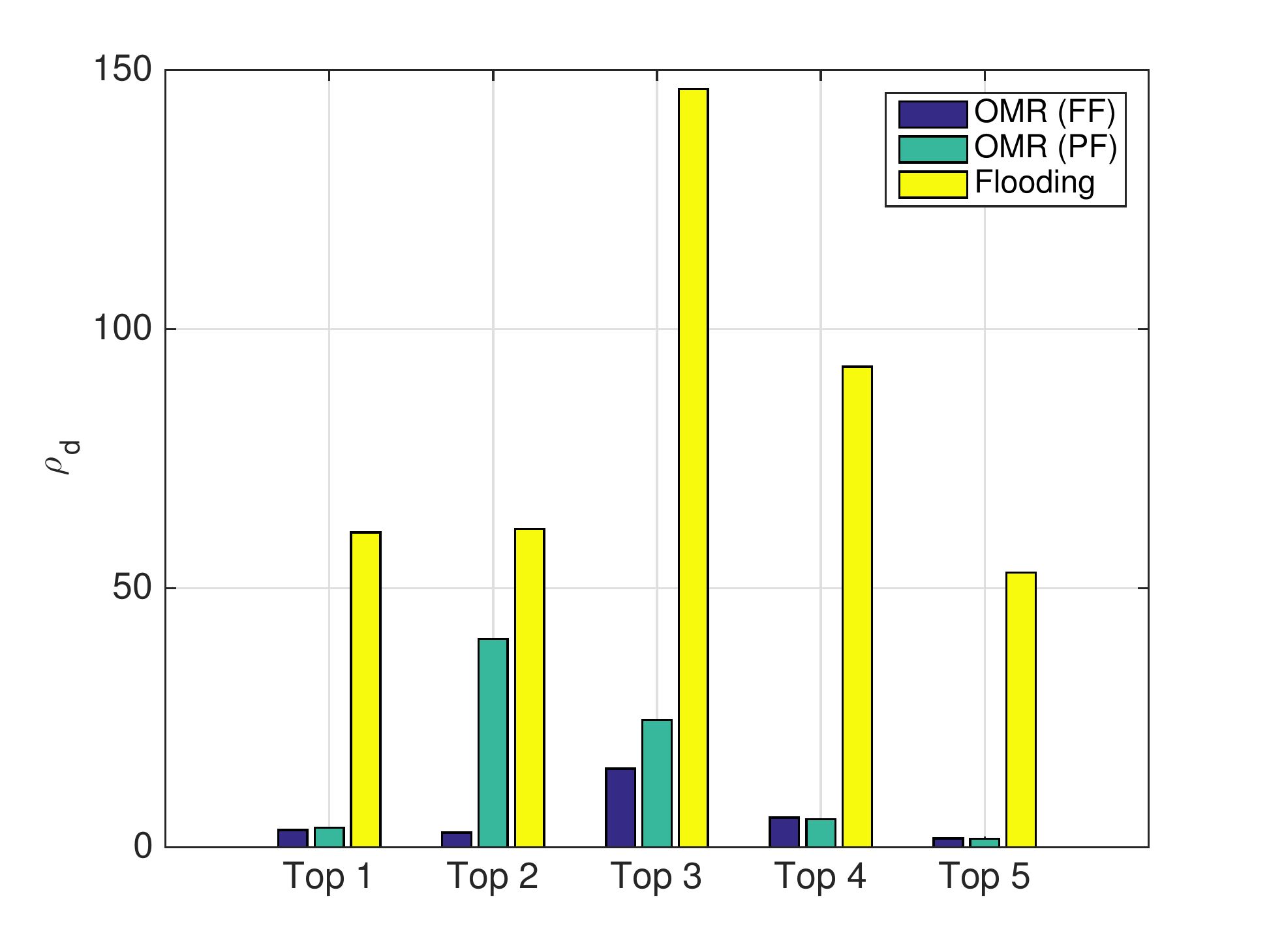}}
    \subfloat[$\rho_g$ (see Eq.~\eqref{e:goodput})\label{f:ExpGoodput}]{\includegraphics[width=0.4\textwidth]{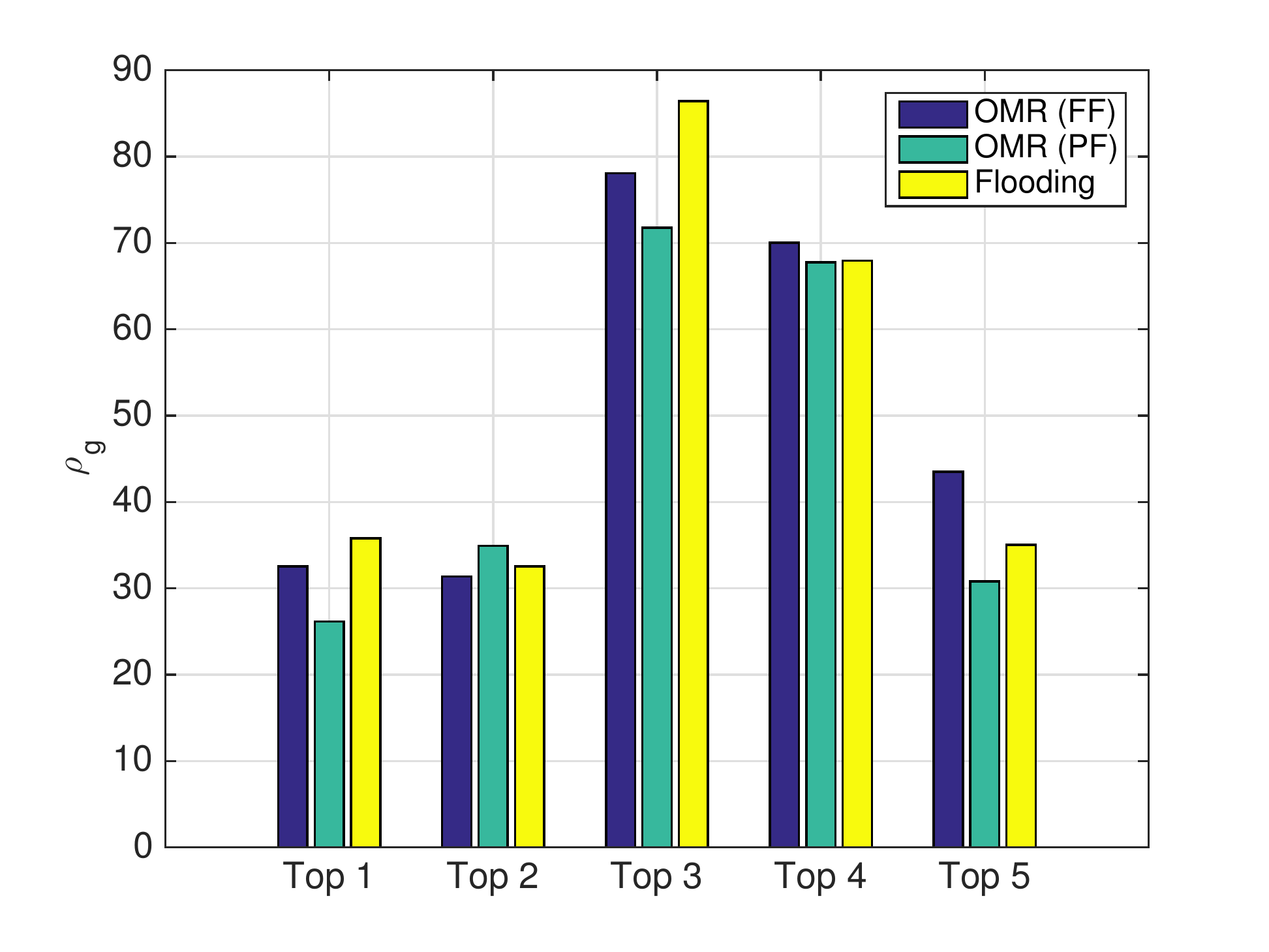}}
    \caption{Experiment: $\rho_d$ and $\rho_g$. flooding performs even worse than in simulation; OMR--FF performs better than OMR--PF.}
\end{figure}

In Fig.~\ref{f:ExpDelay}, we show the measured end-to-end delay $\rho_d$ (see~\eqref{e:delay}) for flooding and OMR in each topology. We observe that the end-to-end delay of flooding is significant, due to the many collisions (and subsequent re-transmissions) caused by the forwarding of every packet over every available technology. On the contrary, the OMR methods performs quite similar to the simulations, with OMR--FF achieving better results than OMR--PF. 
Still, in some cases OMR--PF achieved shorter transmission delay than OMR--FF. This is because OMR--PF in general uses more links than OMR--FF, which tends to be an advantage in the presence of many collisions. 

Fig.~\ref{f:ExpGoodput} shows the goodput results (see $\rho_g$ from~\eqref{e:goodput}). We observe that due to the higher number of packet collisions in a real environment, flooding's goodput decreased compared to the simulations, becoming similar and sometimes lower than that of OMR--FF. Due to the use of full topology information, the Goodput of OMR--FF is higher than that of OMR--PF. An exception to the latter result is seen in Topology~2: the reason is that this topology offers many similar routes from each node to the sink, and thus spreading the transmissions over multiple links has a positive effect.

\begin{figure}[t]
    \centering
    \subfloat[LF links.\label{f:LinkUtilization_low}]{\includegraphics[width=0.32\textwidth]{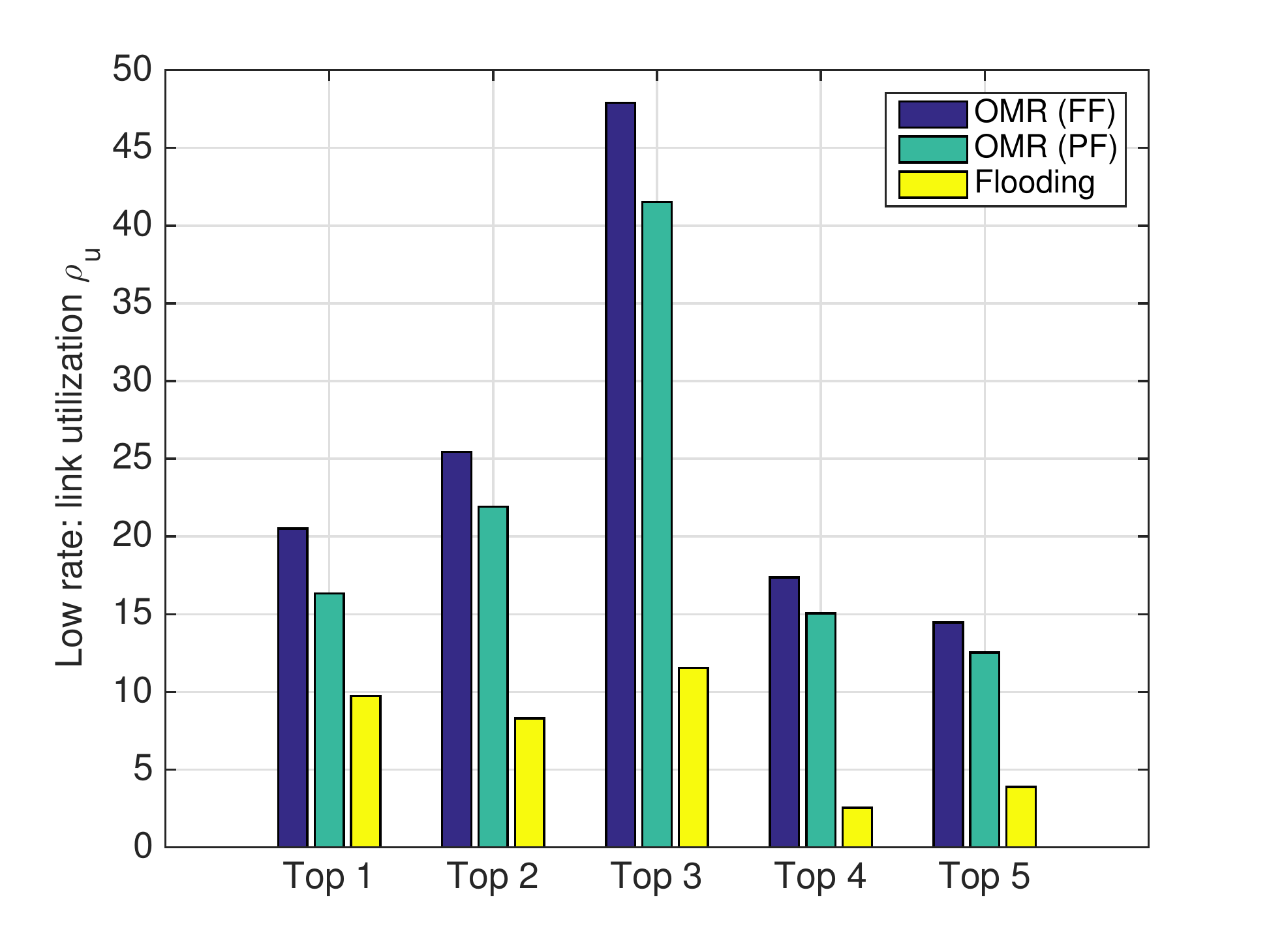}}
    \subfloat[MF links.\label{f:LinkUtilization_mid}]{\includegraphics[width=0.32\textwidth]{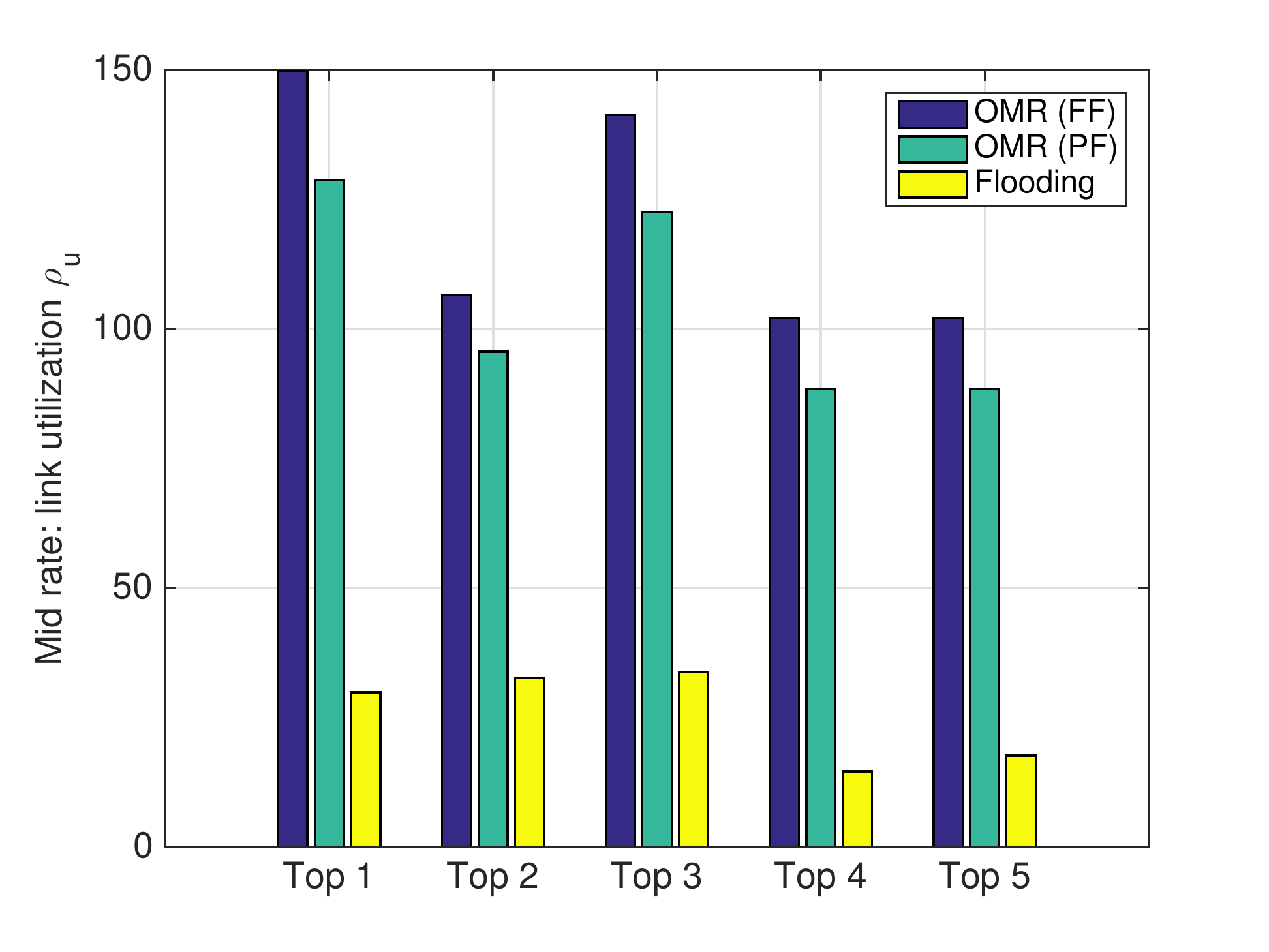}}
    \subfloat[HF links.\label{f:LinkUtilization_high}]{\includegraphics[width=0.32\textwidth]{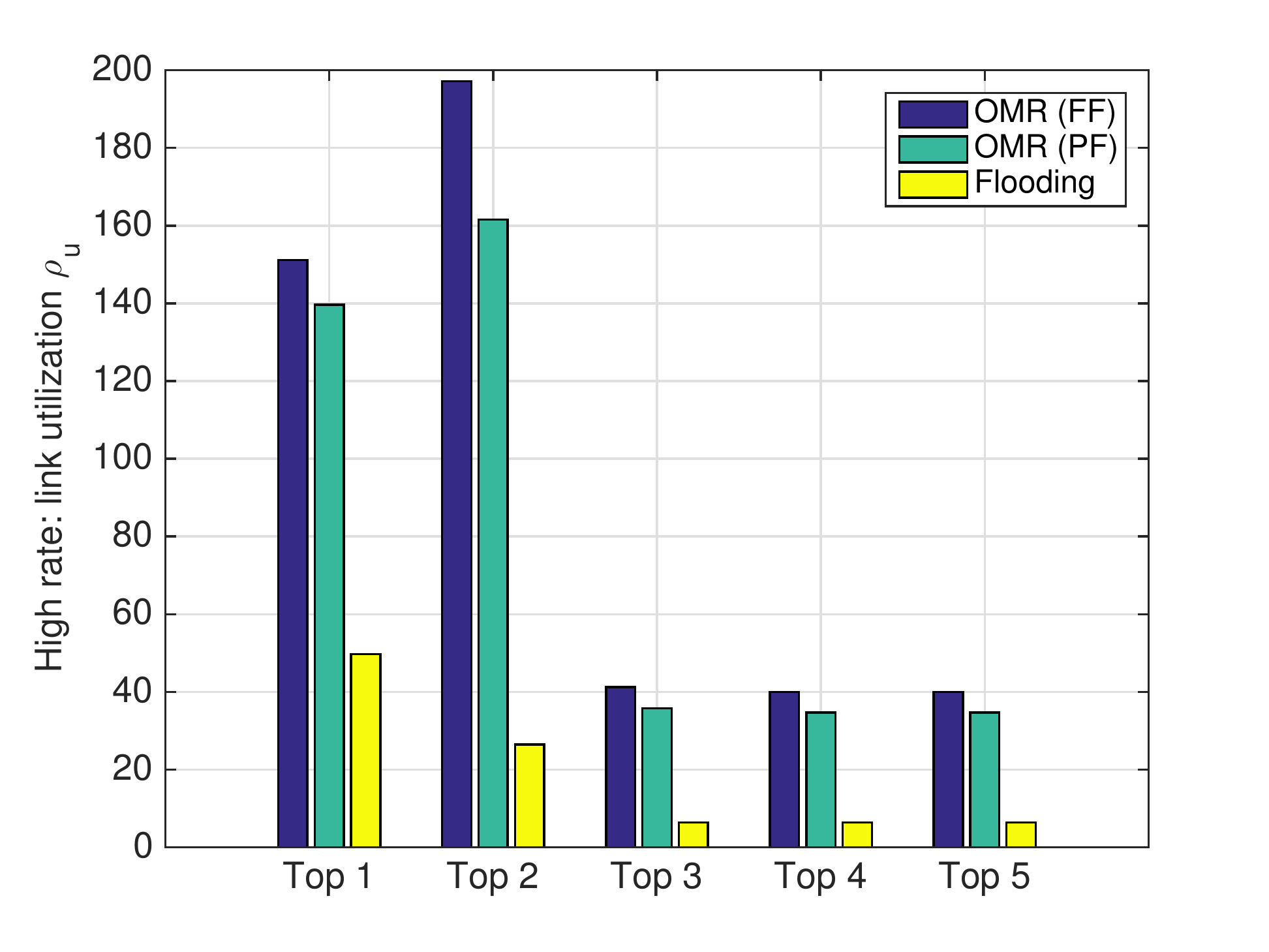}}
    \caption{Experiment: $\rho_u$. flooding performs worse compared to the simulations. OMR--FF achieves better results than OMR--PF.}
    \label{f:LinkUtilization}
\end{figure}

The per-topology link throughput $\rho_u$, (see~\eqref{e:LU}), is shown in Fig.~\ref{f:LinkUtilization} for the three simultaneously used communication types. For all communication technologies, we observe that OMR--FF delivers the best performance and that, although flooding produces many more transmissions over each link, the link throughput of the two OMR methods is significantly higher. 
Again, this is a consequence of the many packet collisions that occur. For the same reason, the link throughput of OMR--FF is better than that of OMR--PF. Comparing the link throughput for the three communication types, we observe that OMR channels more transmissions through links with higher capacity. As a result, the network adapts itself to the topology, as confirmed by the changes in the link throughput for the five topologies tested in the experiment, each having a different configuration of multi-modal links.

%

%% file: conclusions.tex
\section{Conclusions}
\label{sec:concl}

In this paper, we considered the network operation of multi-modal systems, a technology which holds great benefit to underwater networks. We proposed OMR that, to the best of our knowledge, is the first optimal routing protocol to be specifically designed for multi-modal underwater networks and to be experimented in the field.
Our protocol leverages either full or local topology knowledge to decide how to distribute traffic over available links using available communication technologies. This is achieved in a way that does not congest the relays upstream, and reserves more link resources for the nodes with fewer routing opportunities. We analyzed the performance of OMR by means of both simulations and field experiments. Our results show that, even in the presence of imperfect topology information, our protocol leverages the available technologies (and, if available, topology information) to deliver data reliably without congesting the network. 

%% file: appendix_disjoint_routes.tex
\section{Disjoint route computation algorithm in the presence of complete topology information}

We now introduce the algorithm employed to identify all possible disjoint routes in case full topology information is available to a node. This means that the node is fully aware of the topology graph ${\cal G} = ({\cal N},{\cal E})$, where ${\cal N}$ is the set of network nodes and any directed edge $(\ell,m) \in {\cal E}$ represents an existing link between nodes $\ell,m \in {\cal N}$. 

The pseudo-code of the algorithm to find disjoint routes between node $i$ and the destination $D$ is provided in Algorithm~\ref{alg:disj_pathfinder}. The approach is based on a modified version of the Max-Flow algorithm~\cite[Ch.~6]{book_alg_maxflow}. 
Starting from ${\cal G}$, node $i$ constructs a different graph ${\cal G}' = ({\cal N}',{\cal E}')$, where each $n \in {\cal N}$ is substituted by two nodes $n_{\rm in}, n_{\rm out}$ in ${\cal N}'$ (line~\ref{line:disj_gprime2}), where the former accepts incoming links, and the latter emanates outgoing links. The nodes $n_{\rm in}$ and $n_{\rm out}$ are connected by a directed link from the former to the latter in ${\cal E}'$ (line~\ref{line:disj_gprime3}). For every link $(\ell,m) \in {\cal E}$, the set ${\cal E}'$ is further populated with one directed link that connects node $\ell_r{\rm out}$ to node $m_{\rm in}$ (line~\ref{line:disj_eprime2}).
In summary, ${\cal N}' = \{ n_{\rm in}, n_{\rm out} | n\in {\cal N} \}$, and ${\cal E}' = \{(\ell_{\rm out},m_{\rm in}) | (\ell,m) \in {\cal E} \} \cup \{ (n_{\rm in}, n_{\rm out}) | n \in {\cal N} \}$. Finally, node $i$ assigns a unit weight to all links in ${\cal E}'$ (line~\ref{line:disj_wprime2}), and solves a maximum flow problem between $i$ and $D$ over ${\cal G}'$ with weights $W$, where $W_{ij}$ is the weight assigned to the directed link $(i,j)$ (line~\ref{line:disj_route_finder}). The solution of the problem is equal to the number of node-disjoint routes from $i$ to $D$, denoted as $\PCL{i}{j}$ in Section~\ref{sec:routing}. 

\begin{algorithm}[t]
\label{Algo1}
    \renewcommand{\baselinestretch}{1}\small
    \caption{Disjoint route search algorithm}\label{alg:disj_pathfinder}
    \DontPrintSemicolon
    \SetKwProg{Fn}{Function}{}{end}
    \SetKwInput{KwRequire}{Require}
    \Fn{\textsc{FindDisjointRoutes}\,(\,\,${\cal G} = ({\cal N},{\cal E}), \, i, \, D$\,\,) }{
        ${\cal N}' \gets \varnothing, \; {\cal E}' \gets \varnothing$   \label{line:disj_init}   \;
        \tcp{Generate ${\cal G}' = ({\cal N}',{\cal E}',W')$}
        \ForEach{$n \in {\cal N}$    \label{line:disj_gprime1}   }{
            ${\cal N}' \gets {\cal N}' \cup \{n_{\rm in}, n_{\rm out}\}$    \label{line:disj_gprime2}   \;
            ${\cal E}' \gets {\cal E}' \cup \{ \, (n_{\rm in}, n_{\rm out}) \, \}$    \label{line:disj_gprime3}   \;
        }
        \ForEach{$(\ell,m) \in {\cal E}$    \label{line:disj_eprime1}   }{
            ${\cal E}' \gets {\cal E}' \cup \{ \, (\ell_{\rm out}, m_{\rm in}) \, \}$    \label{line:disj_eprime2}   \;
        }
        \ForEach{$(\ell,m) \in {\cal E}$    \label{line:disj_wprime1}   }{
            $W'_{\ell m} \gets 1$    \label{line:disj_wprime2}   \;
        }
        ${\cal R} = \textsc{MaxFlow}( {\cal N}',\, {\cal E}',\, W' ,\, i_{\rm out},\, D_{\rm in})$   \label{line:disj_route_finder}   \;
    }
\end{algorithm}
